\documentclass[12pt]{iopart}
\usepackage[
backend=biber,
]{biblatex}
\usepackage{graphicx}
\usepackage{bm}
\usepackage{mathrsfs}
\usepackage{amsmath}
\usepackage{floatrow}

\usepackage{amssymb}
\usepackage{bbold}
\usepackage{appendix}
\usepackage{xcolor}
\usepackage{float}
\usepackage{braket}
\usepackage{comment}
\usepackage{diagbox}
\allowdisplaybreaks
\DeclareMathOperator\sign{sign}

\begin{document}
\title{Interacting Dreaming Neural Networks}

\author{Pietro Zanin and Nestor Caticha}
 
\address{
 Instituto de  Fisica, Universidade de  Sao Paulo\\ Sao Paulo, SP, Brasil
}%
\ead{{10739927@usp.br}, {ncaticha@usp.br}}
\vspace{10pt}
\begin{indented}
\item[]\today 
\end{indented}

\begin{abstract}
We study the interaction of agents, where each  one consists of an associative memory neural network trained with the same memory patterns and possibly different reinforcement-unlearning dreaming periods. Using replica methods, we obtain the rich equilibrium phase diagram of the coupled agents. It  shows phases such as the student-professor phase, where only one network benefits from the interaction while the other is unaffected; a mutualism phase, where both benefit; an indifferent phase and an insufficient phase, where neither are benefited nor impaired;  a phase of amensalism where one is unchanged and the other is damaged.  In addition to the paramagnetic and spin glass phases,  there is also one we call the reinforced delusion phase,  where agents concur without having finite overlaps with memory patterns. For zero coupling constant, the model becomes the reinforcement and removal dreaming model, which without dreaming is the Hopfield model. For finite coupling and a single memory pattern, it becomes a Mattis version of the Ashkin-Teller model. 


\vspace{1cm}
The authors declare that they have no conflict of interest.

\vspace{1cm}
Partial Funding from:  CNPq National Council for Scientific and Technological Development and Fapesp, the São Paulo Research Foundation.

\vspace{.5cm}
'This is the Accepted Manuscript version of an article accepted for publication in Journal of Statistical Mechanics: Theory and Experiment: J. Stat. Mech. (2023) 043401. Neither SISSA Medialab Srl nor IOP Publishing Ltd is responsible for any errors or omissions in this version of the manuscript or any version derived from it.  The Version of Record is available online at : http://dx.doi.org/10.1088/1742-5468/acc72b
\end{abstract}
\maketitle

\section{\label{sec:level} Introduction}
Neural networks are supposed to implement meaningful information processing based on data. Meaningful, of course, has to be defined. We are interested in the behavior of interacting neural networks, where the data available to each network is partially provided by  other neural networks. Shared meaning emerges from this interaction. We obtain analytical results, for a system of two interacting agents, each one modeled by an associative memory.  Each agent receives the same memory patterns from the environment, however they may undergo different levels of post-learning dynamics, in the form of {\it dreaming},  i.e. unlearning or reinforcing by relearning. 
Hopfield, Feinstein and Palmer \cite{Hopfield1983}, following ideas from  Crick and Mitchinson \cite{CrickMitchinson1983}, introduced a learning algorithm based on unlearning that improved significantly the performance of the Hopfield \cite{Pastur, Hopfield82} network with pure Hebbian learning. This anti-Hebbian mechanism was inspired by the suggestion  \cite{CrickMitchinson1983} that rapid eye moving (REM) type dreaming, permits mammalians to forget spurious memories, increasing their capacity to retrieve relevant memories. Further analysis and simulations by van Hemmen {\it et al.}  \cite{Hemmen1992} and \cite{WimabuerKlemmerHemmen1994}, showed that iterating this algorithm initially increases the network’s capacity. However, if iterated beyond a  certain threshold, retrieval is destroyed. It was shown that this increment occurs in part due to the fact that the synaptic matrix approaches the pseudo-inverse synaptic matrix \cite{KanterSompolinsky1987}, despite not converging to it. Dotsenko \cite{Dotsenko1991} used a similar algorithm to justify a new symmetric synaptic matrix, that allowed reaching the theoretical upper bound in the thermodynamic limit: number of patterns $p$, equal to the  number of units $N$, but only at $T=0$. 

Fachechi {\it et al.} \cite{Fachechi2019} showed analytically that beneficial unlearning plus the enhancement of the pure states, which they called  reinforcement,  can increase the stability of the patterns, increasing the critical capacity. Their model has a much larger retrieval phase than the Hopfield model, not only saturating the maximal capacity at $T=0$ but also increasing the stability of the model to temperature changes.  Most importantly, there is no catastrophic upper bound on the length of the dreaming process. This algorithm resembles developments by Plakhov, Semenov and Shuvalova \cite{SemenovShuvalova95}  \cite{PlakhovSemenov94}, but results in a more interesting and varied synaptic matrix. Also relevant for this discussion are \cite{Barra2012},\cite{Alemanno2022},\cite{Fachechi2022}, where they manage to connect effectively learning in Hopfield models with learning in Boltzmann machines.

Dreaming, which is a process internal to the neural network, represents an extra study period of the available information by the agent. We allow agents to interact through a coupling that 
extends this internal dreaming to an externally driven process, that also acts on the minima of the Hamiltonian. We show that this process has the  potential to make the networks perform  better or worse, depending on the conditions of the system.

\section{Model of interacting neural networks}

\subsection{Introduction of the model}

We work on a statistical mechanics problem in the canonical formalism with temperature $T= \beta^{-1}$. The components of the quenched memory patterns $\bm \xi =\{\xi_i^\mu \}_{\mu=1\cdots p, i=1\cdots N} $, are drawn independently from a uniform Bernoulli distribution. The Hamiltonian of an isolated agent is \cite{Fachechi2019}
\begin{eqnarray}
\mathcal{H}_{1}(t,\bm \xi, \bm \sigma)&=&-\frac{1}{2N}\sum_{i,j=1}^N\sum_{\mu,\nu=1}^p\xi_{i}^{\mu }\xi_{j}^{\nu }(\frac{1+t}{\bm 1+t\bm C})_{\mu\nu }\sigma_{i}\sigma_{j},
\end{eqnarray}
where  $t\ge 0$ quantifies the length of the dreaming process. $\bm C$ is the $p\times p$ matrix with elements $C_{\mu \nu}=\frac{1}{N}\sum_{i=1}^{N}\xi_{i}^{\mu}\xi_{i}^{\nu}$, $\sigma_{i}=\pm 1$ is an Ising variable. For  $t=0$ the  Hopfield Hebbian Hamiltonian is recovered and as $t\rightarrow\infty $, $t^{-1}$  acts as a Tikhonov regularizer, leading to the pseudo-inverse model \cite{Personnaz85}\cite{KanterSompolinsky1987}.
The two member group Hamiltonian is 
\begin{eqnarray} 
\mathcal{H}(t_1,t_2,\bm \sigma, \bm S,\epsilon) &=&
\mathcal{H}_{1}(t_1,\bm \xi, \bm \sigma)+\mathcal{H}_{1}(t_2,\bm \xi, \bm S)\nonumber+\frac{\epsilon}{N}(\sum_{i}\sigma_{i}S_{i})^2,
\label{hamilton}
\end{eqnarray} 
 $\epsilon$ is the coupling constant of the interaction, the sums over $i,j$ run from $1$ to $N$, the sums over $\mu,\nu$ from $1$ to $p$. We study this problem in equilibrium or offline with quenched disorder. A related problem, the inverse problem of learning online a spin glass Hamiltonian,  was studied in \cite{KuvaKinouchiCaticha97}. 

The first two terms of the Hamiltonian in  expression \ref{hamilton} represent two distinct neural network agents, the third term connects them. 

This interaction is similar to dreaming  since it changes the couplings from $J^{(1)}_{ij}\rightarrow J_{ij}^{(1)}+\frac{2\epsilon}{N}S_{i}S_{j}$ and $J^{(2)}_{ij}\rightarrow J_{ij}^{(2)}+\frac{2\epsilon}{N}\sigma_{i}\sigma_{j}$, i.e. they pick up a Hebbian contribution from the other agent. However, we emphasize the fact that they are different, as in the interaction all states are included, and not only the stables ones, as it is done in the unlearning algorithm.

For $\epsilon=0$, this is the model of reinforcement and removal model \cite{Fachechi2019}; and with $\epsilon \neq 0$ and $p=1$, the infinite range Askhin-Teller model\cite{KadanoffWegner71}.

Without any loss, we take $t_1 \le t_2$ and refer to the neural network agents as NN1 and NN2 respectively.

\subsection{Order parameters}
The free-energy is obtained using the replica method, where the Hamiltonian is a sum of the individual replica Hamiltonians. The calculation  follows roughly those for the  single agent case by Fachechi {\it et al.}  \cite{Fachechi2019}, except for a few important steps where it is necessary to be careful.  The expression of the free energy and the self-consistent equations for the order parameters appear  in \cite{SuppMat}.
Since we are interested  in the memory retrieval phases, we have not looked in detail into the spin-glass phase, hence, in order to understand global features of the five dimensional phase diagram  ($t_{1}, t_{2},\epsilon, \alpha, \beta$), the important order parameters for replica $a$ are $h^{a}=\langle\langle\frac{1}{N}\sum_{i=1}^{N}\sigma_{i}^{a}S_{i}^{a}\rangle\rangle$, $m_{1}^{a}=\langle\langle\frac{1}{N}\sum_{i=1}^{N}\xi^{1}_{i}\sigma_{i}^{a}\rangle\rangle$ and $m_{2}^{a}=\langle\langle\frac{1}{N}\sum_{i=1}^{N}\xi_{i}^{1}S_{i}^{a}\rangle\rangle$, where the double brackets represent the thermal and quenched  average over the patterns. We will also  consider  the changes induced by the interaction, i.e. the difference between the order parameters for $\epsilon> 0$ and $\epsilon=0$, for a more careful identification of the different phases.
Other order parameters of the form $\Delta^{\rho}$, $r^{\rho}$, $Q^{\rho}$ and $q^{\rho}$ are related to the overlap between the $\rho$ elements, where $\rho$ is an index that refers to either agent $\sigma$, or    $S$ or to their overlap   $\sigma S$.

We assume the following replica symmetry ansatz: 

\begin{eqnarray}
\nonumber
\\&&m_{1}^{a}=m_{1};\,\,\, m_{2}^{a}=m_{2};\,\,\, h^{a}=h\,\,\,\, \forall a,  \label{ansatz}
\\&&q_{ab}^{\rho}=Q^{\rho}\delta_{ab}+q^{\rho}(1-\delta_{ab})\,\,\,\,\forall a,b,\rho\nonumber
\\&&r_{ab}^{\rho}=R^{\rho}\delta_{ab}+r^{\rho}(1-\delta_{ab})\,\,\,\, \forall a,b,\rho.\nonumber
\end{eqnarray}
where
\begin{eqnarray}
q_{ab}^{\rho}=\langle\langle\frac{1}{N}\sum_{i}(\sigma_{i}^{a,\rho}+i\sqrt{\frac{t}{\beta(1+t)}}\phi^{a,\rho})(\sigma_{i}^{b,\rho}+i\sqrt{\frac{t}{\beta(1+t)}}\phi^{b,\rho})\rangle\rangle,
\end{eqnarray}
and $r_{ab}$ is the auxiliary variable of $q_{ab}$, so it has a similar physical meaning.

The first line of (\ref{ansatz}) is not hard to justify, since there are no reasons to expect that any  replica  is privileged. 

The Ansatz related to $q_{ab}^{\rho}$ and $r_{ab}^{\rho}$ for  $\rho=\sigma S$ deserves a comment.  It represents the idea that the correlations between the agents will be different for the same replica than for different replicas, because the interaction is not between different replicas, but within the same, so we expect a greater intrareplica similarity of the agents than the interreplica similarity. 

A homogeneous Ansatz ($q^{\sigma S}_{ab}=q^{\sigma S}, r^{\sigma S}_{ab}=r^{\sigma S}\forall a,b$), leads to the puzzling result that the free energy does not depend on $q^{\sigma S}$, which indicates that this is not the right Ansatz. This can be seen from the fact that the dependence of $Q^{\sigma S}$ and $q^{\sigma S}$ on the free energy occurs via the differences $Q^{\sigma S}-q^{\sigma S}$ and $R^{\sigma S}-r^{\sigma S}$, so using a homogeneous Ansatz would lead to no dependence on $q^{\sigma S}$. 

It is important to recognize that $q_{ab}$ does not have the exact same meaning as the original Edwards-Anderson parameter. Its additional complex part inside the average leads to different values, and in particular, $q^{\sigma S}$ is different from $h$ for $t_{1},t_{2}\neq 0$, despite having some similarity. It also explains why already in the original article \cite{Fachechi2019} the usual choice $q_{aa}=1$ was not used. Still, this parameter gives us a measure of the correlations in the system, and can still be used to determine the existence of the spin-glass state.

We emphasize that despite using the same Ansatz for different $\rho$'s, the justification behind them is distinct. The replica symmetric free energy depends on 9 independent order parameters ($m_{1}, m_{2}$, $h$, $Q^{\sigma}$, $Q^{S}$, $Q^{\sigma S}$, $q^{\sigma}$, $q^{S}$, $q^{\sigma S}$). We also have 6 dependent variables ($r^{\sigma}$, $r^{S}$, $r^{\sigma S}$, $\Delta^{\sigma}$, $\Delta^{S}$, $\Delta^{\sigma S}$), which can be fully eliminated from the equations of state.

Table \ref{table:phases} shows the different phases, obtained from the possible regimes of $m_1, m_2, h$ and, in addition, a measure of the benefit or damage due to the interaction, given by 
\begin{eqnarray}
&&\Delta m_1= m_1(\epsilon,t_{1},t_{2},\alpha,\beta) - m_1(0,t_{1},t_{2},\alpha,\beta),
\\&&\Delta m_2= m_2(\epsilon,t_{1},t_{2},\alpha,\beta) - m_2(0,t_{1},t_{2},\alpha,\beta),\nonumber
\\&&\Delta h= h(\epsilon,t_{1},t_{2},\alpha,\beta) - h(0,t_{1},t_{2},\alpha,\beta)\nonumber
\end{eqnarray}

\begin{table}[ht]
\begin{tabular}{ |p{3.5cm}||p{.8cm}|p{.8cm}|p{.8cm}|p{.8cm}|p{.8cm}|p{.8cm}|  }
 \hline 
 \diagbox[width=3.2cm]{Phases}{{OP\,\,\,\,\,\,\,\,\,}} & $m_{1}$ & $m_{2}$ & $h$ & $\Delta m_{1}$ & $\Delta m_{2}$ & $\Delta h$\\
 \hline\hline 
 
 Student-professor &   $> 0$  & $> 0$   & $> 0$ & $> 0$ & $\approx 0$ & $> 0$\\
 Mutualism &   $> 0$  & $> 0$   & $> 0$& $> 0$ & $>0$ & $> 0$\\
 Disordered    &   $\approx 0$ & $\approx 0$   & $\approx 0$& $\approx 0$ & $\approx 0$ & $\approx 0$\\
 Reinforced delusion&   $\approx 0$ & $\approx 0$   & $> 0$& $\approx 0$ & $\approx 0$ & $> 0$\\
 Insufficient   & $\approx 0$    & $>$ 0 & $ \approx 0$& $\approx 0$ & $\approx 0$ & $\approx 0$\\
 Indifferent &   $> 0$  & $> 0$   & $> 0$& $\approx 0$ & $\approx 0$ & $\approx 0$\\
 Amensalism &   $\approx 0$  & $\approx 0$   & $\approx 0$& $\approx 0$ & $< 0$ & $\approx 0$\\
 \hline
 \end{tabular}
\caption{The different phases are characterized by the values of the order parameters (OP), we are considering that $t_{1}\leq t_{2}$ and that $m_{1},m_{2},h\geq 0$. }
\label{table:phases}
 \end{table}
 
Before entering into details in section \ref{Phasediagrams}, we give a rough description of the types of phases where the system may be found.  The {\it insufficient} phase occurs when NN2 is in the retrieval phase and NN1 is not and $\epsilon$ is small enough such that  no relevant changes in the learning occurs. As the interaction increases, it can  transition to the {\it student-professor} phase, where the performance of only NN1 is enhanced significantly. Further increase of $\epsilon$ can result in the {\it mutualism} phase, where both neural networks have noticeable increases in their capacity.
The {\it disordered} phase includes the paramagnetic and spin glass phases, where the interaction cannot lead to interesting results. It is not hard to detect the spin glass phase, since for $h=0$, the role of $\epsilon$ vanishes, hence the line of the spin glass phase is simply the same obtained in \cite{Agliari2019}.\\

The {\it indifferent} phase occurs where, even without interaction, both neural networks could already process information in an adequate manner. The {\it reinforced delusion}  phase indicates the creation of an alternate ordered state, where the agents, despite having  no overlap with the memory patterns, are similar to each other.  From their own perspective, the agents  are in a retrieval ferromagnetic phase.
For a third party, the individual agents cannot process information, and seem to be in a disordered phase. The interaction still preserves the glassy state, so it is still a spin glass phase. 
The last phase is the {\it amensalism}, where while NN1 is not significantly affected by the interaction, NN2 ends up losing the understanding of the situation, replicating an interaction of amensalism.

\subsection{Phase diagrams and discussion \label{Phasediagrams}}

We analyze numerically the self-consistent equations obtained from the extreme conditions of the free energy. Three types of  2-dimensional cuts are used to describe the five dimensional parameter space, 
yielding phase diagrams in the: (i) $\alpha, T_{c}$, (ii) $\epsilon, \alpha_{c}$ and (iii)  $\epsilon, T_{c}$ planes, shown respectively in panels \ref{painel1}, \ref{painel2} and \ref{painel3}. In all diagrams, since $t_{2}\geq t_{1}$, we have $m_{2}\geq m_{1}$.

\subsubsection{ $T$-$\alpha_{c}$ plane}

\begin{figure}[ht]
  \includegraphics[width=0.45\textwidth]{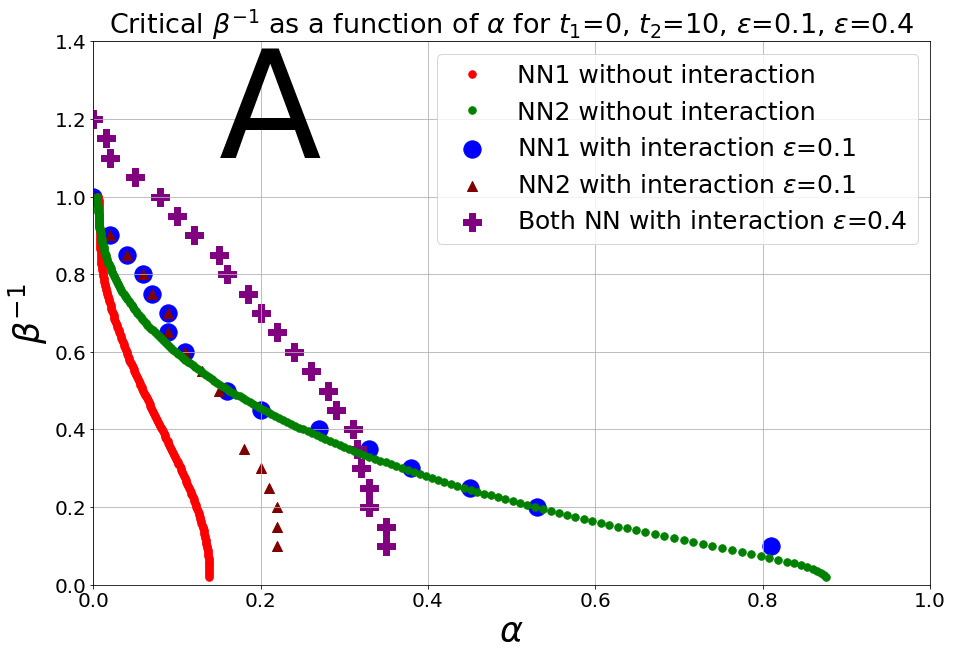}%
  \includegraphics[width=0.45\textwidth]{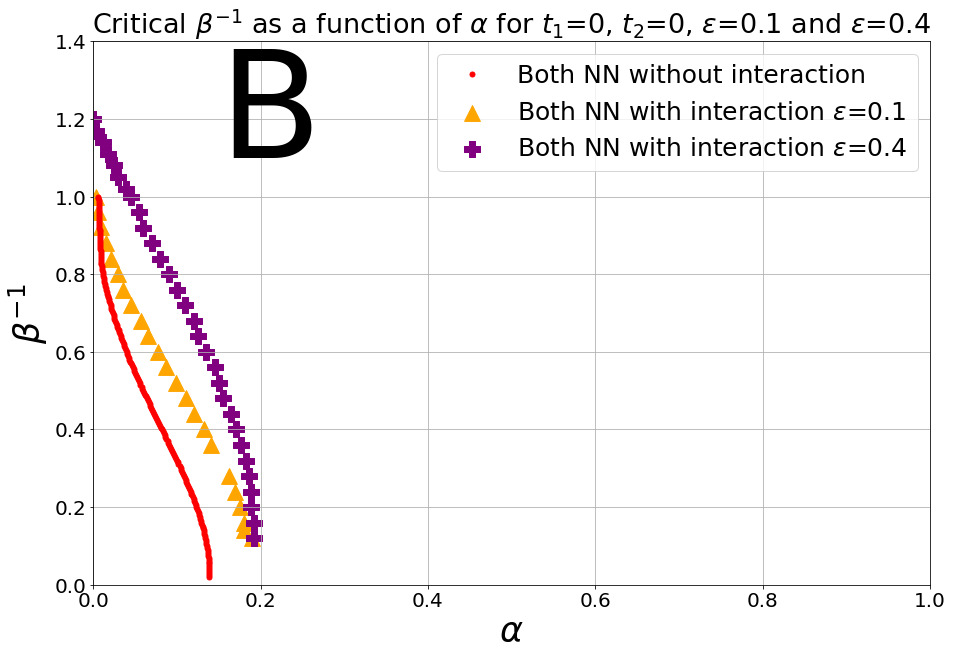}%
\\
  \includegraphics[width=0.45\textwidth]{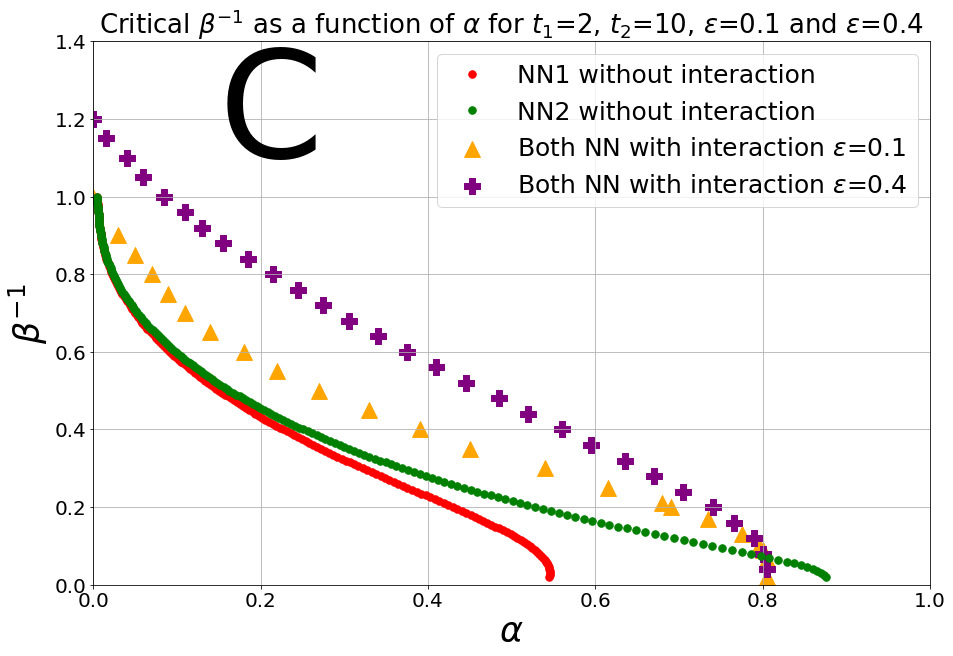}
  \includegraphics[width=0.45\textwidth]{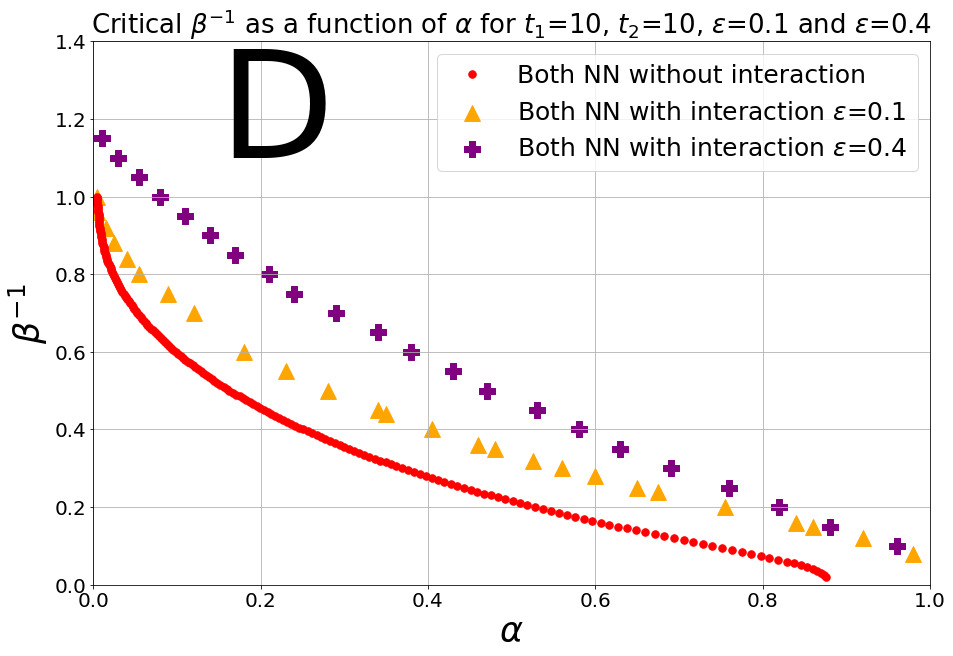}
\caption{
The metastable  phase border  critical capacities $\alpha_{c,i}(T; t_1,t_2, \epsilon)$. 
A) Symmetrical case with $t_{1}=t_{2}=0$ and $\epsilon=0,0.1,0.4$. Ferromagnetic (FM) phase increases with the interaction and no dreaming. Note that for high $\epsilon$, at $\alpha =0$, the FM phase persists up to $T_c>1.$
B)  $t_{1}=0=t_{2}=0$ and $\epsilon=0,0.1,0.4$. For low temperatures,  high interactions lead to a NN2's critical capacity decrease, while it is always beneficial for the critical capacity of NN1.
C) $t_{1}=2$, $t_{2}=10$ and $\epsilon=0,0.1,0.4$. A similar to (B), but with $t_{1}$ higher and closer to $t_{2}$, even for small $\epsilon$ the two agents have the same behavior. Loss of capacity for the NN2 is much smaller here, since NN1 is not harmful.
D) $t_{1}=t_{2}=10$ for $\epsilon=0,0.1,0.4$. Interaction between well-trained peers is helpful.
 }\label{painel1}
 \end{figure}

For fixed $t_{1},t_{2}$ and $\epsilon$, $T$ is  varied to find the maximum capacity such that $m_{1}\neq 0$ or $m_{2}\neq 0$. We compare them to the non-interacting neural networks. The main results can be seen in  panel \ref{painel1}.


The behavior can be understood by comparing the difference between the properties of an interacting agent with $t_{1}$ or $t_{2}$, with the properties of the non-interacting agents with the same amount of dreaming. 

Diagram \ref{painel1}.B shows the no dreaming interacting agents, $t_1=t_2=0$. The interaction yields a higher critical capacity and both agents benefit equally from the interaction.
Diagram \ref{painel1}.A represents the typical situation of having a high value of the dreaming disparity $t_1=0, t_2=10$. Agent 1 always benefits, since its capacity is always larger than without interaction. For small interaction, $\epsilon =0.1$, agent 2 slightly benefits for low $\alpha$ but is damaged by the interaction at higher values of $\alpha$. For high values of the  interaction, say  $\epsilon=0.4$,  they have the same value of critical capacity for all temperatures. Agent 2 benefits considerably from the interaction at low $\alpha$, but is again damaged at higher loads, because it is learning from the spurious minima of the low dreaming agent 1, which is able to benefit only up to a certain point by the interaction.


Diagram \ref{painel1}.C represents the typical situation of having an intermediate and high dreaming load, $t_1=2, t_2=10$. Above a certain $\epsilon<0.1$,  $\alpha_{c,1}=\alpha_{c,2}$  and in general the interaction is substantially beneficial for both  neural networks, except at very low temperatures $T\lesssim 0.05$, where there is a slight decrease in the pattern retrieval critical capacity of agent 2. 

Diagram \ref{painel1}.D presents the case where both agents have dreamt abundantly and equally  $t_1=t_2=10$. Again $m_{1}=m_{2}$ by symmetry, and the interaction increases significantly the phase of the pattern retrieval, and differently from the previous situation it is always beneficial. As in diagram \ref{painel1}.A, interacting with a peer, brings an advantage to both agents, independently of dream load. 

\subsubsection{$\epsilon$-$\alpha_{c}$ plane}

\begin{figure}[ht]
\centering{%
  \includegraphics[width=0.40\textwidth]{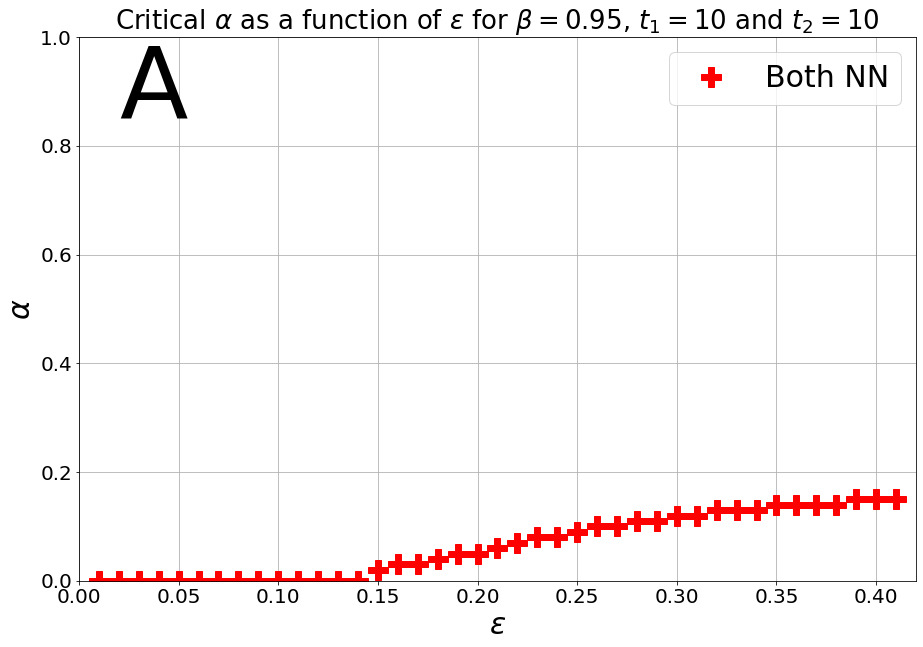}%
}\qquad
{%
  \hspace*{0.1em}
  \includegraphics[width=0.40\textwidth]{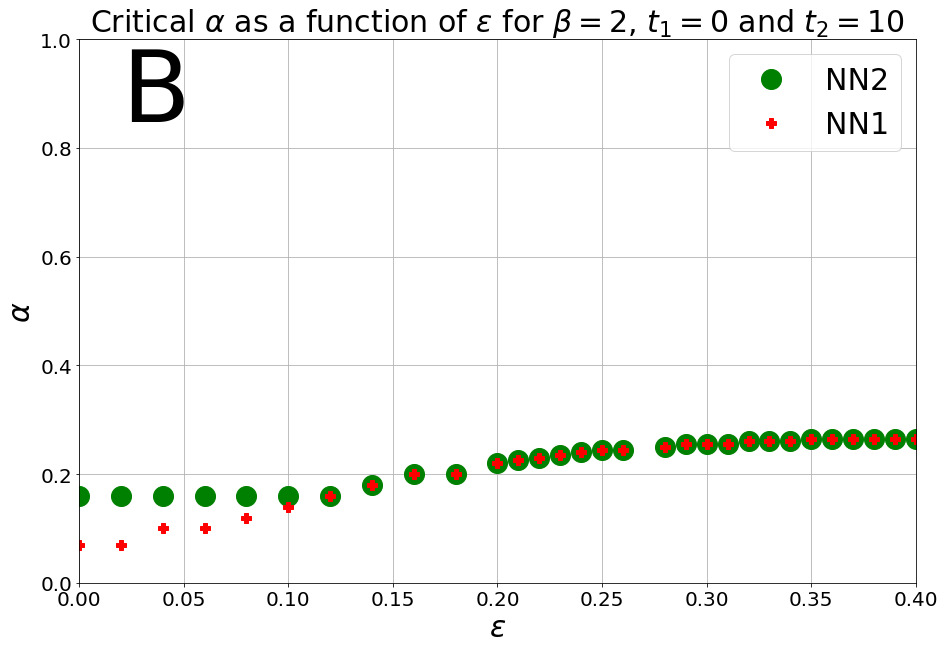}%
}\\
\hfill \break
\centering
{%
  \includegraphics[width=0.40\textwidth]{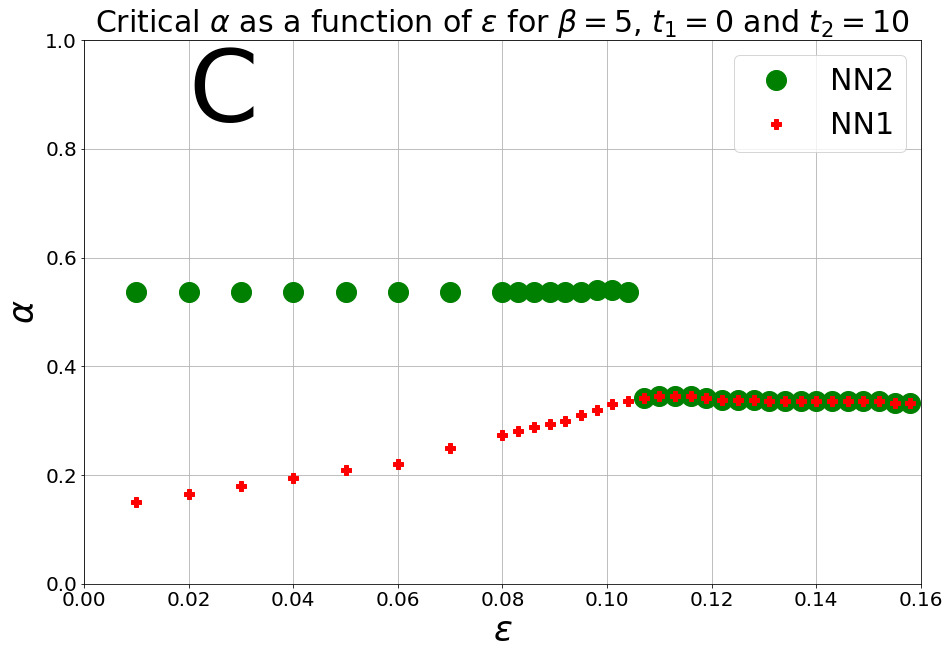}%
}\qquad
{%
  \hspace*{0.1em}
  \includegraphics[width=0.40\textwidth]{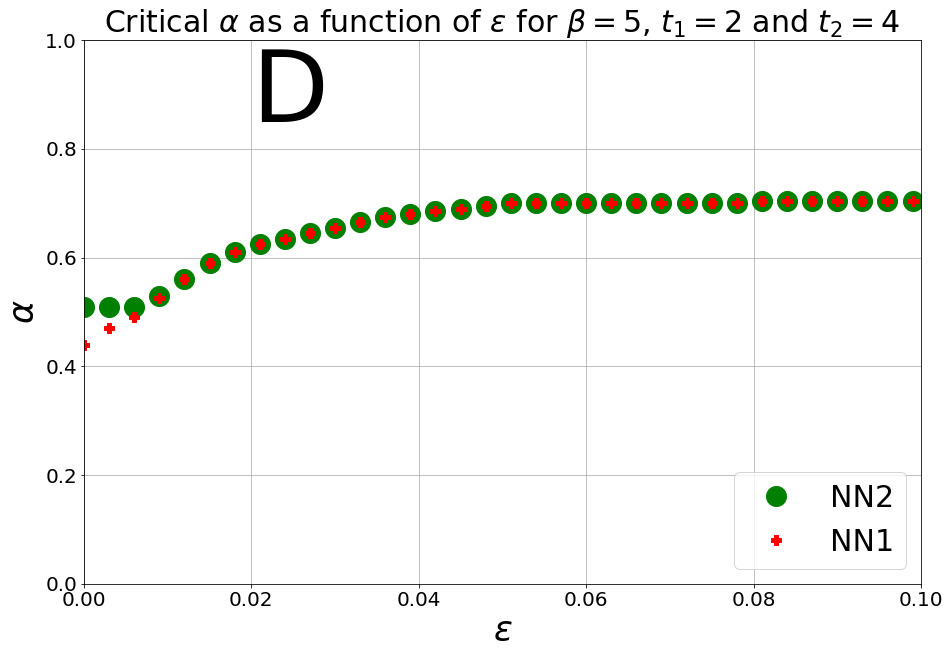}%
}
\caption{Critical capacities $\alpha_{c,i}(\epsilon; t_1,t_2, T)$ .
A) $\beta=0.95$ and $t_{1}=t_{2}=10$, there is only one critical line since $t_{1}=t_{2}$; we see a typical situation of mutualism, where the two agents manage to obtain non-zero magnetization with strong interaction. 
B) $\beta=2$, $t_{1}=0$, $t_{2}=10$, typical student-professor situation for small interaction, the critical capacity of NN2 is constant up to $\epsilon\approx 0.12$ and then turns to mutualism.
C) $\beta=5$, $t_{1}=0$ and $t_{2}=10$, the student-professor phase turns to amensalism, with NN2 harmed by the interaction.
D)$\beta=5$, $t_{1}=2$ and $t_{2}=4$. A  student professor situation at low $\epsilon$ turns to mutualism for larger $\epsilon$. } \label{painel2}
\end{figure}

For $t_{1},t_{2}$ and $T$ fixed, we  varied $\epsilon$ and $\alpha$ to obtain the value of $\alpha_{c,i}(\epsilon)$ such that for all $\alpha>\alpha_{c,i}(\epsilon)$ we have $m_{i}=0$. In particular, we have significant differences at low temperatures $(T\lesssim 0.3)$ and high temperatures ($0.3\lesssim T$). The main results can be seen in panel \ref{painel2}.

The first 2 diagrams \ref{painel2}.A and \ref{painel2}.B represent the high temperature region. While the second diagram shows the mutual benefits of the interaction for networks with high difference of $t$, the first shows that with enough interaction, it is possible to  extend the capacity to regions where it was impossible before the interaction. This retrieval above  $T=1$ occurs because the coupled system acts as a system with a higher number of neurons. An interesting detail shown in diagram \ref{painel2}.B is that for low values of the interaction, only agent 1 has an increase in the capacity, while agent 2 is unchanged. It only modifies its retrieval capacity when the capacities start matching. The former student reaches the level of the professor and both can profit from the interaction. 

Panels \ref{painel2}.C and \ref{painel2}.D show the behavior for the low temperature region $T=0.2$. Diagram \ref{painel2}.D shows that for low differences of $t$ we still have a qualitatively similar behavior as in diagram \ref{painel2}.B. However, for big differences in dreaming load, there is a discontinuous transition at high values of the interaction in which the capacity of agent 2 decreases substantially. It is easy to see that this behavior is consistent with diagram \ref{painel1}.B. Of course, the critical capacity of agent 2, despite having no benefit from the interaction, is still larger than in the high temperature case shown in figure \ref{painel2}.B.
For the parameters where the two agents are beneficial to each other, the increase in their interaction is not detrimental. We see this in the increase of the upper boundary of the yellow region. But this cannot improve forever, and the alphas (figure \ref{painel2}.A, \ref{painel2}.B and \ref{painel2}.D) tend to a limit value.

\subsubsection{$\epsilon$-$T_{c}$ plane}

Here, we fixed $t_{1},t_{2},T$ and varied $\epsilon$ and $T$ to obtain the value of $T_{c}(\epsilon)$ such that for $T>T_{c}(\epsilon)$ we have $m_{1}=0$ or $m_{2}=0$. In these, differently from the previous diagrams, we incorporated the denomination of different phases mentioned in subsection II.B, as they are easier to visualize. The main results can be seen in  panel \ref{painel3}:

\begin{figure}[ht]
\centering{
  \includegraphics[width=0.40\textwidth]{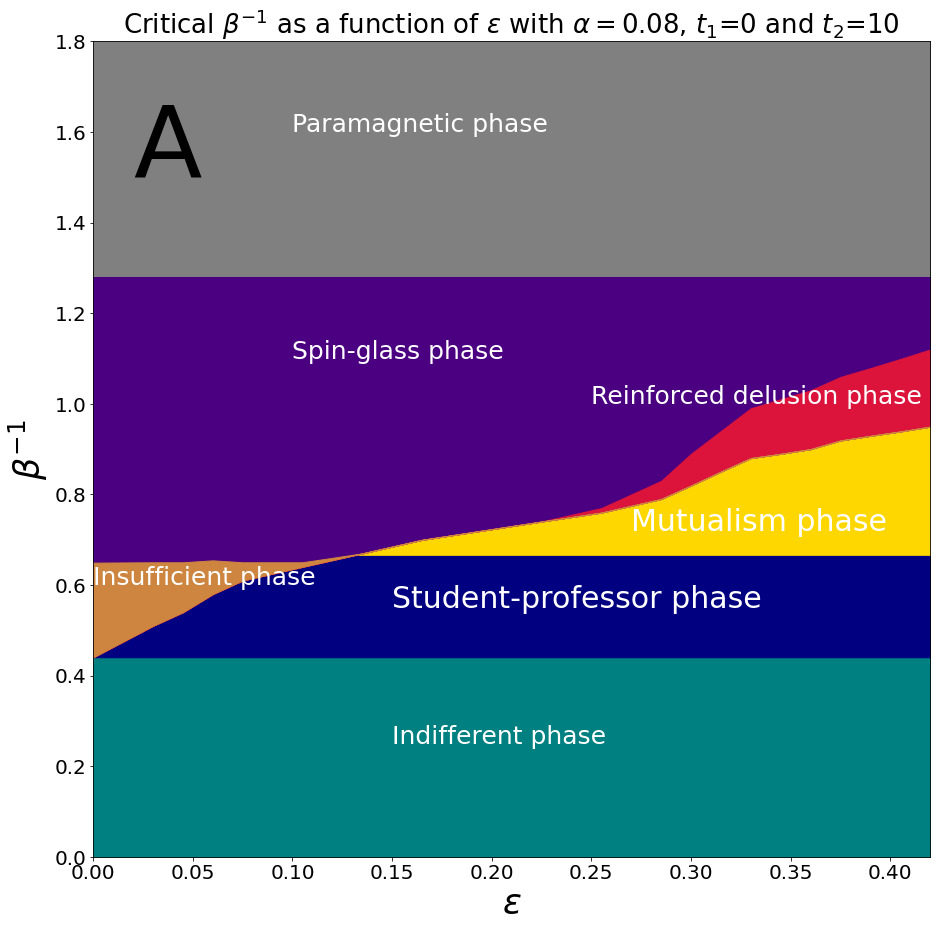}
}\qquad
{%
  \hspace*{0.1em}
  \includegraphics[width=0.40\textwidth]{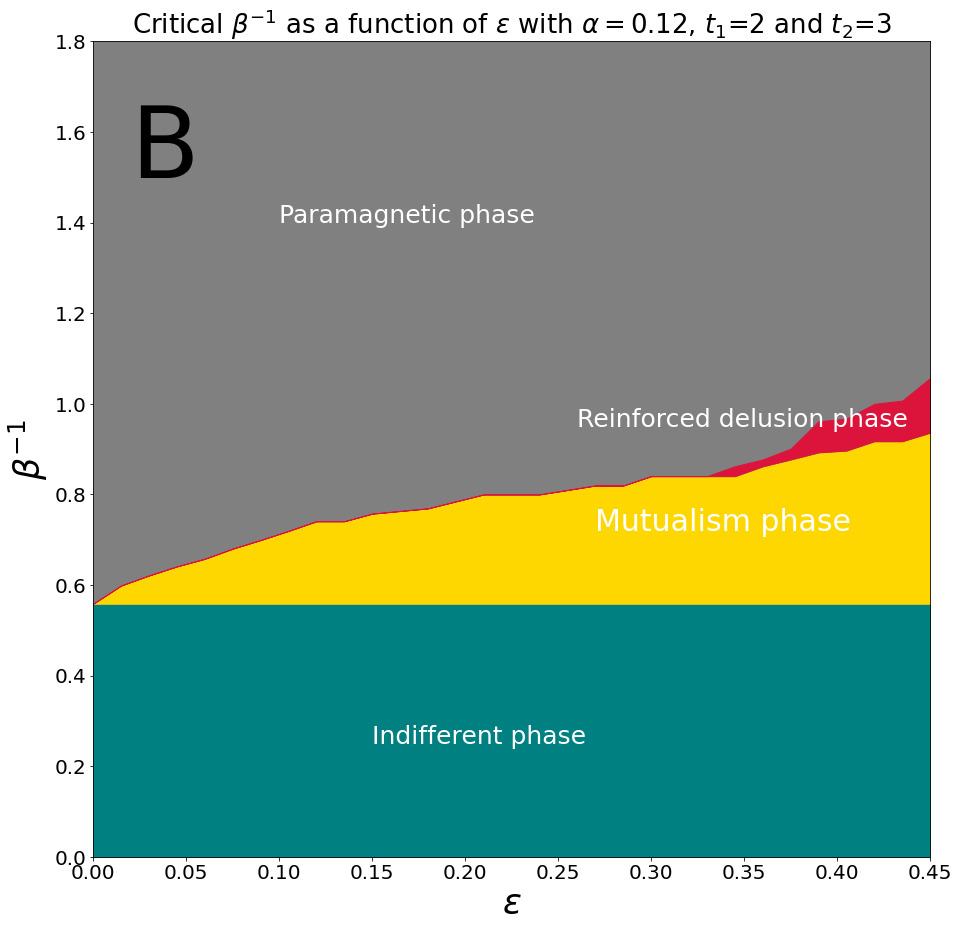}%
}\\
\hfill \break
\centering
{%
  \includegraphics[width=0.40\textwidth]{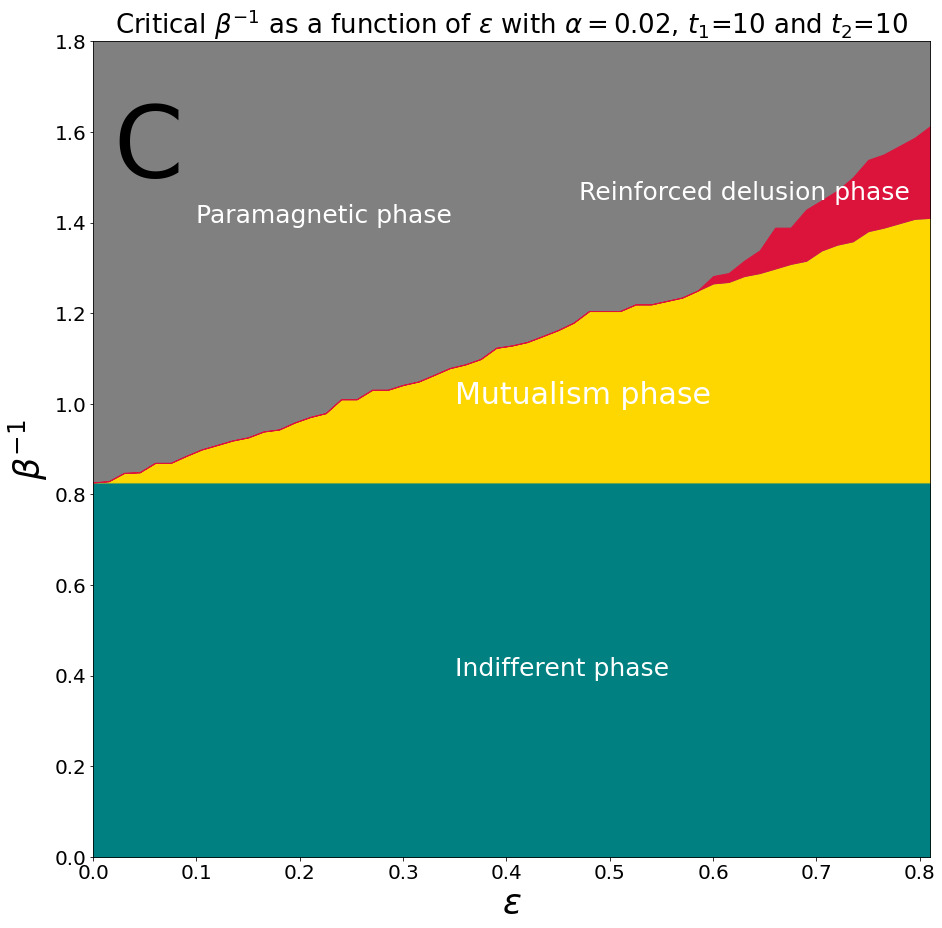}%
}\qquad
{%
  \hspace*{0.1em}
  \includegraphics[width=0.40\textwidth]{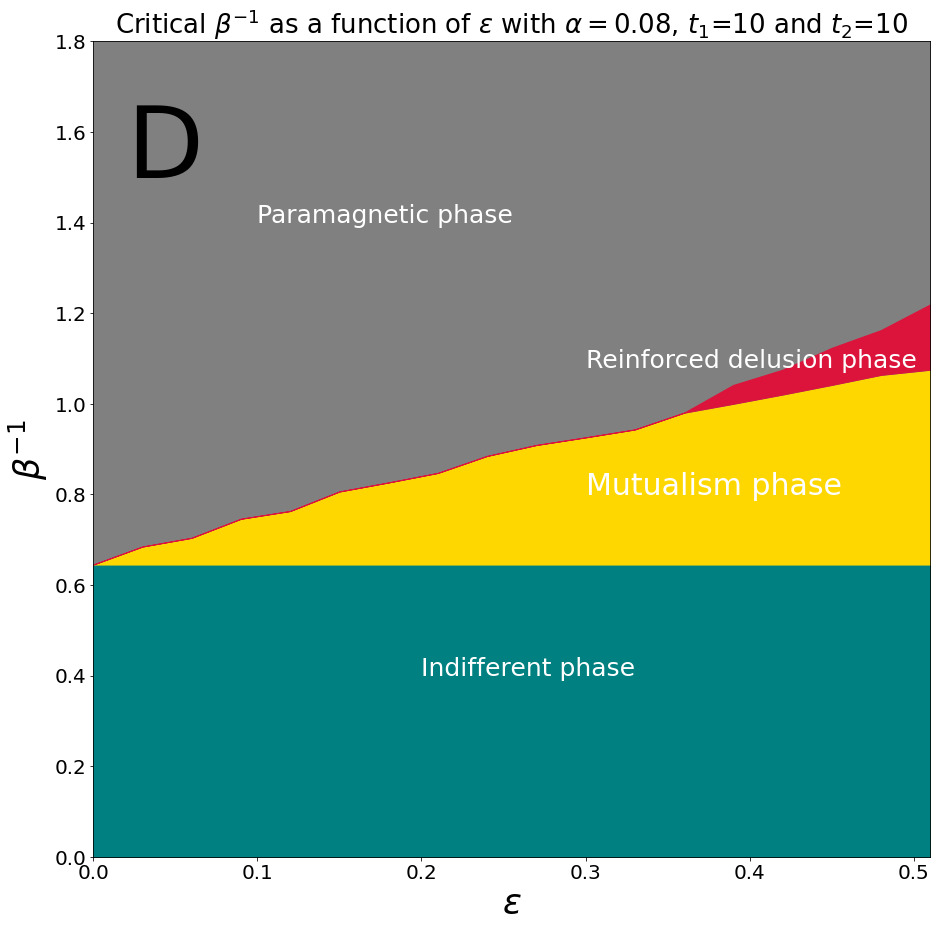}}%
\caption{Phase diagram in the temperature $T,\epsilon$ plane for fixed$ t_1,t_2, \alpha)$.  
A) For fixed $\alpha=0.08$, $t_{1}=0$ and $t_{2}=10$. Since the difference between $t_{1}$ and $t_{2}$ is large, both the insufficient phase and the student-professor phase are present. 
B) With $\alpha=0.12$, $t_{1}=2$ and $t_{2}=3$, the student-professor and the insufficient phases have disappeared due to a small difference between $t_{1}$ and $t_{2}$. 
C)  $\alpha=0.02$ and $t_{1}=t_{2}=10$. Same dreaming loads leads to symmetry and therefore no  student-professor nor insufficient phases. This also happens in 
D)  $\alpha=0.08$ and $t_{1}=t_{2}=10$, note that the disordered and reinforced delusion phases increase and the mutualism phase decreases significantly.}\label{painel3}
\end{figure}

The indifferent phase is usually small and only appears in models where the $t_{1}$ and $t_{2}$ difference is high, in our examples it can only be visualized in the first case. The mutualism phase only increases as the interaction increases, and tends to a plateau for high values, as expected. The reinforced delusion only appears for high values of interaction and increases beyond the mutualism phase plateau. It is important to explain that the behavior visualized in diagram \ref{painel3}.A is not in contradiction with diagram \ref{painel2}.C, as in the diagram 3.A, we have a low  $\alpha$ value, so the phenomenon of decreasing magnetization with interaction does not occur. Additionally, we do not see the amensalism phase in panel \ref{painel3} because we are only considering relatively low capacities, where this phase is not present.

\begin{figure}[t]
\centering{%
  \includegraphics[width=0.40\textwidth]{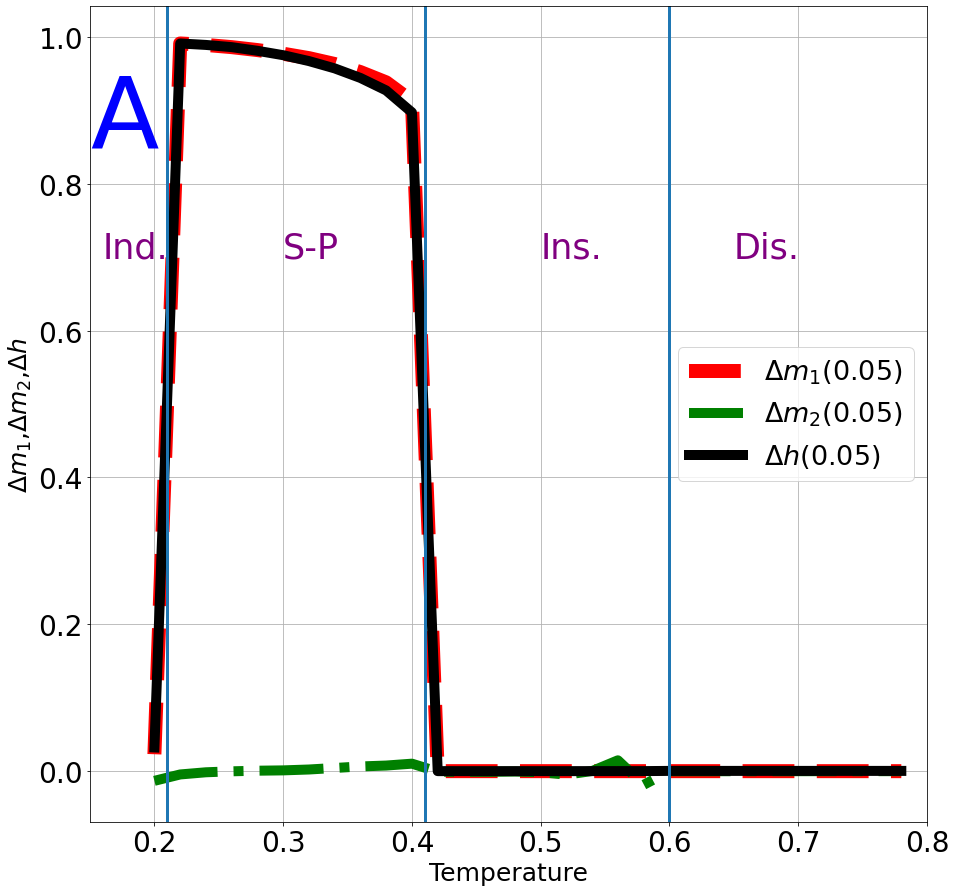}%
}\qquad
{%
  \hspace*{0.1em}
  \includegraphics[width=0.40\textwidth]{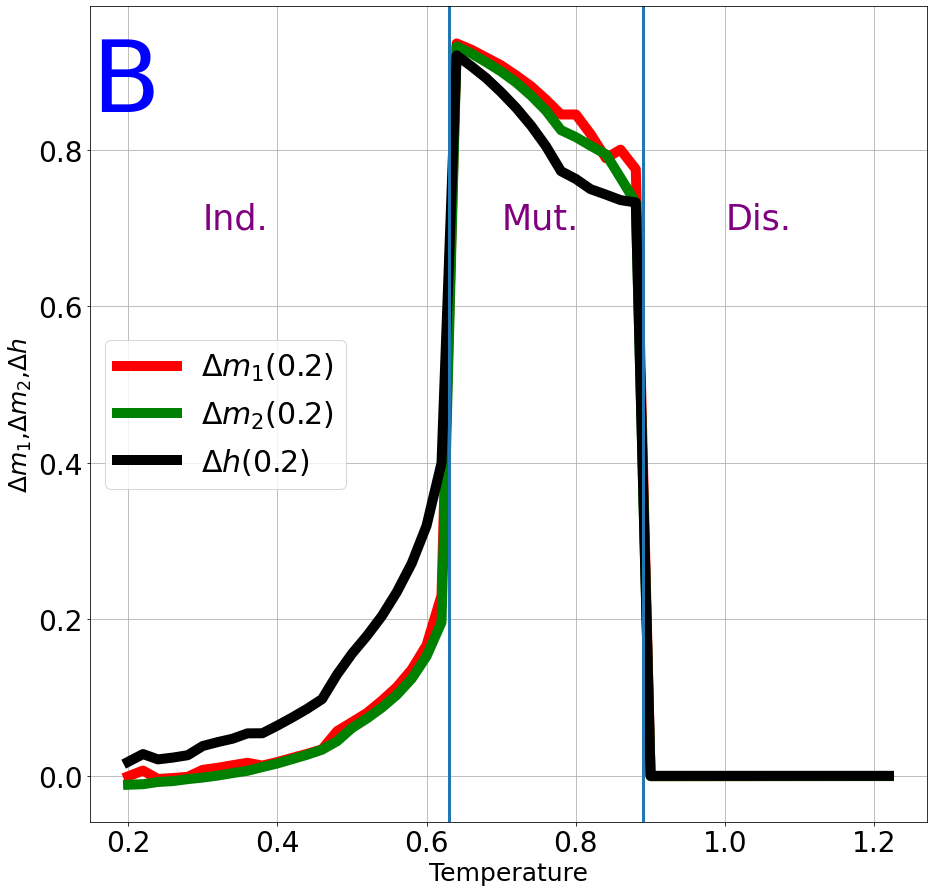}%
}\\
\hfill \break
\centering
{%
  \includegraphics[width=0.40\textwidth]{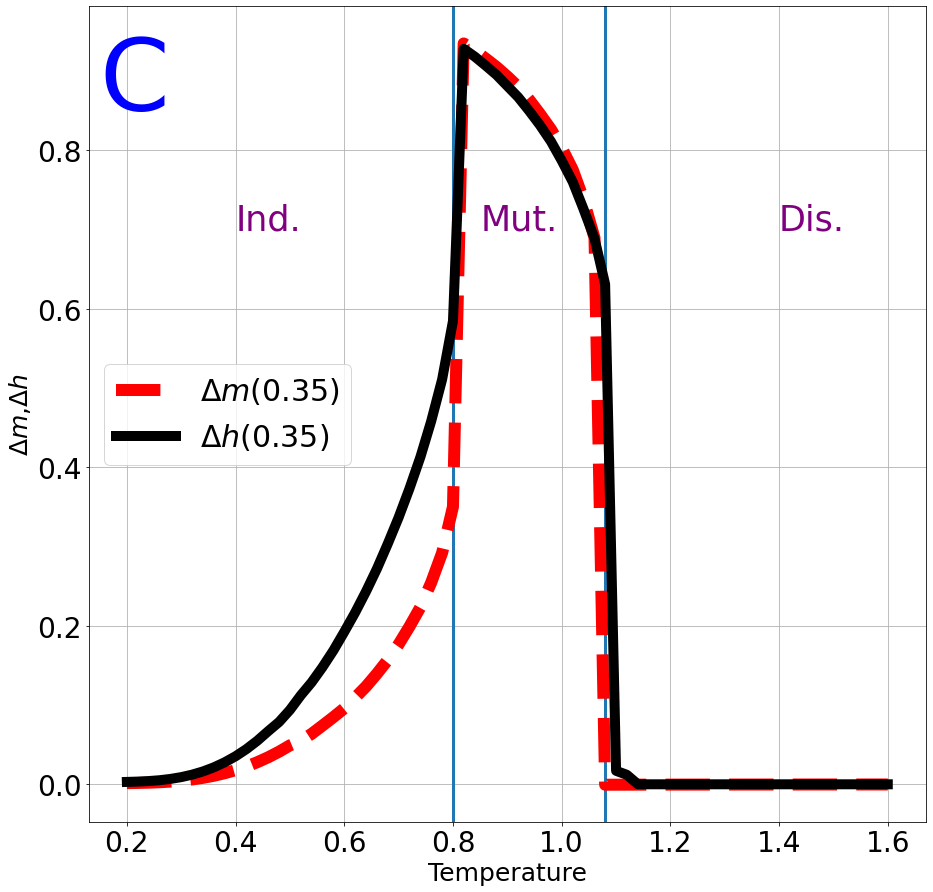}%
}\qquad
{%
  \hspace*{0.1em}
  \includegraphics[width=0.40\textwidth]{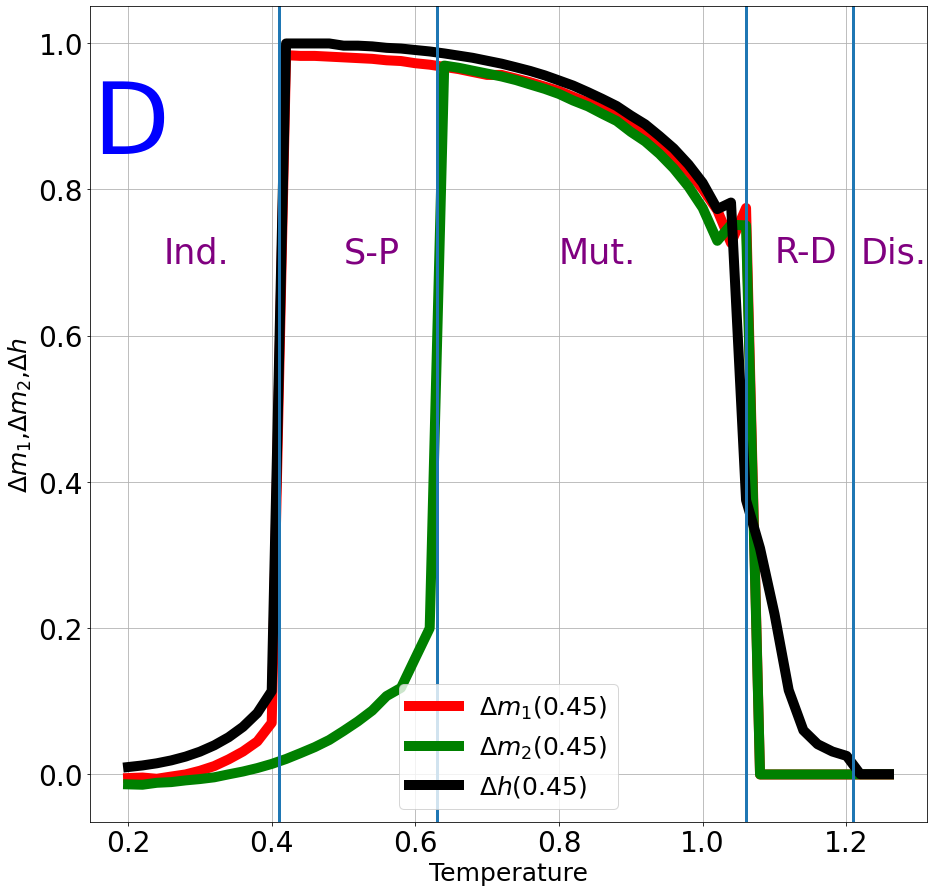}%
  
}
\caption{ $\Delta m_{i}(T; t_1,t_2, \alpha,\epsilon)$ and $\Delta h(T; t_1,t_2, \alpha,\epsilon)$. The legend shows $\Delta m, \Delta h$ for a particular cut of $\epsilon$ given $(t_1,t_2, \alpha)$.
A) A cut with $\alpha=0.12$, $\epsilon=0.05$, $t_{1}=0$ and $t_{2}=10$; it is interesting to compare it with diagram 3.A, as they are similar situations. Disordered phase here corresponds to the spin-glass phase.
B) A cut with  $\alpha=0.08$, $\epsilon=0.2$, $t_{1}=2$ and $t_{2}=3$; compare it to diagram 3.B. Disordered phase here corresponds to the paramagnetic phase.
C)  $\alpha=0.02$, $\epsilon=0.35$ and $t_{1}=t_{2}=10$, we write $m\equiv m_{1}=m_{2}$ as they are equal, compare it to the diagram 3.C and 3.D. Disordered phase here corresponds to the paramagnetic phase.
D) With $\alpha=0.08$, $\epsilon=0.45$, $t_{1}=0$ and $t_{2}=10$, see that the small tail in the $\Delta h$ at high values of $\epsilon$ indicates the reinforced delusion phase; compare this image to diagram 3.A. Disordered phase here corresponds to the spin-glass phase.}\label{painel4}
\end{figure}

To get a more clear view of the phase changes in this model, we add panel \ref{painel4} that shows how $\Delta m_{1}$, $\Delta m_{2}$ and $\Delta h$ change with temperature for fixed $\epsilon$, $\alpha$, $t_{1}$ and $t_{2}$

\section{Conclusions}
Statistical Mechanics techniques were used to model and study the interaction  between information processing machines. 
Simple exposure to information, without the benefit of post-processing, in the form of dreaming, is not efficient.
The agents received the same information in the form of $P= \alpha N$ patterns.
The removal and reinforcement of minima can be thought of as a further  exercise, of duration $t_1$ and $t_2$, that the agents  undergo after being exposed to the information in the memory patterns; in a metaphorical sense, the agents ponder, meditate, think over, dream about the received information. In addition to the individual learning process, further changes in their properties are elicited by an interaction quantified by $\epsilon$, and a rich and non-trivial behavior ensues; interactions can be irrelevant,  beneficial or harmful to the information retrieval. We return to what is meaningful information processing? It depends on who assigns meaning. If the retrieval is used to gauge the success of the machines, as an independent third party would, the region of {\it reinforced delusion} shows no meaningful information processing. However, from the perspective of the agents, it seems to be fine. They agree on their version of what is correct, despite it doesn't reflect the objectively relevant information in the memory patterns.
The ubiquitous use of machines that learn demands the study of their interactions, not only with other machines, but with humans too. Modifications, generalizations or simplifications of our approach are needed. 


\vspace{1cm}

\ack{We thank Felippe Alves Pereira for useful discussions. This work was partially supported by FAPESP and CNPq. The number of FAPESP's project is nº2021/07951-7.}

\printbibliography

\appendix

{\Large {\bf Supplemental Material}}

{\large
{\bf Interacting Dreaming Neural Networks}}

Pietro Zanin and Nestor Caticha

{\it Instituto de Fisica, Universidade de Sao Paulo}
\section{Details of the solution}

The average over the patterns of the replicated partition function is
\begin{eqnarray}
&&\langle Z^n\rangle=\langle\sum_{\sigma^{n},S^{n}}\exp[\frac{\beta}{2N}\sum_{a}\sum_{i,j}\sum_{\mu,\nu}\xi_{i}^{\mu }\xi_{i}^{\nu}(\frac{1+t_{1}}{\mathbb{1}+t_{1}C})_{\mu\nu }\sigma_{i}^{a}\sigma_{j}^{a }
\\&&+\frac{\beta}{2N}\sum_{a}\sum_{i,j}\sum_{\mu,\nu}\xi_{i}^{\mu}\xi_{j}^{\nu}(\frac{1+t_{2}}{\mathbb{1}+t_{2}C})_{\mu\nu}S_{i}^{a}S_{j}^{a}+\beta\epsilon N (\frac{1}{N}\sum_{i}\sigma_{i}^{a}S_{i}^{a})^2]\rangle.\nonumber
\end{eqnarray}

Introducing integrals to remove the inverse matrix and considering that we are only dealing with one condensed pattern ($\bm \xi^1$) :

\begin{eqnarray}
&&\langle Z^n\rangle=\sum_{\sigma^n,S^n}\int\prod_{a}Dx_{1}^{a}Dy_{1}^{a}\prod_{i,a}D\Phi_{i}^{a}D\phi_{i}^{a}\prod_{a}Dh^{a} \\&&\times\exp[\sqrt{\frac{\beta(1+t_{1})}{N}}\sum_{i,a}x_{1}^{a}\xi_{i}^{1}(\sigma_{i}^{a}+i\sqrt{\frac{t_{1}}{\beta(t_{1}+1)}}\phi_{i}^{a})\nonumber
\\&&+\sqrt{\frac{\beta(1+t_{2})}{N}}y_{1}^{a}\xi_{i}^{1}(S_{i}^{a}+i\sqrt{\frac{t_{2}}{\beta(t_{2}+1)}}\Phi_{i}^{a})\nonumber
\\&&+\sqrt{\frac{2\beta\epsilon }{N}}\sum_{a}\sum_{i}\sigma_{i}^{a}S_{i}^{a}h^{a}]\langle\prod_{a}\prod_{\mu\geq 2}Dx_{\mu}^{a}Dy_{\mu}^{a}\exp\{\sum_{i,a}\sum_{\mu\geq 2}\nonumber
\\&&\times\xi_{i}^{\mu}[\sqrt{\frac{\beta(t_{1}+1)}{N}}x_{\mu}^{a}(\sigma_{i}^{a}+i\sqrt{\frac{t_{1}}{\beta(t_{1}+1)}}\phi_{i}^{a})\nonumber
\\&&+\sqrt{\frac{\beta(t_{2}+1)}{N}}y_{\mu}^{a}(S_{i}^{a}+i\sqrt{\frac{t_{2}}{\beta(t_{2}+1)}}\Phi_{i}^{a})]\}\rangle. \nonumber
\end{eqnarray}

The average over the patterns can be done explicitly when we have only one condensed pattern, as it turns out to be a sum of exponentials:

\begin{eqnarray}
&&\langle\exp\{\sum_{ia}\sum_{\mu\geq 2}\xi_{i}^{\mu}[\sqrt{\frac{\beta(t_{1}+1)}{N}}x_{\mu}^{a}(\sigma_{i}^{a}+t_{1}'\phi_{i}^{a})+\sqrt{\frac{\beta(t_{2}+1)}{N}}y_{\mu}^{a}(S_{i}^{a}+t_{2}'\Phi_{i}^{a})]\}\rangle\nonumber
\\&&=\{\cosh[\sum_{ia}\sum_{\mu\geq 2}(\sqrt{\frac{\beta(t_{1}+1)}{N}}x_{\mu}^{a}(\sigma^{a}_{i}+t_{1}'\phi^{a}_{i})+\sqrt{\frac{\beta(t_{2}+1)}{N}}y_{\mu}^{a}(S^{a}_{i}+t_{2}'\Phi^{a}_{i}))]\}^N\nonumber
\\&&=\exp\{\frac{\beta}{2N}\sum_{iab}\sum_{\mu\geq 2}(1+t_{1})(\sigma^{a}_{i}+t_{1}'\phi^{a}_{i})(\sigma^{b}_{i}+t_{1}'\phi^{b}_{i})x_{\mu}^{a}x_{\mu}^{b}
\\&&+(1+t_{2})y_{\mu}^{a}y_{\mu}^{b}(S^{a}_{i}+t_{2}'\Phi^{a}_{i})(S^{b}_{i}+t_{2}'\Phi^{b}_{i})\nonumber
\\&&+2\sqrt{(1+t_{1})(1+t_{2})}[(\sigma^{a}_{i}+t_{1}'\phi^{a}_{i})(S^{b}_{i}+t_{2}'\Phi^{b}_{i})x_{\mu}^{a}y_{\mu}^{b}]\}\nonumber
\\&&=\int\prod_{ab}dq_{ab}^{\sigma}dq_{ab}^{S}dq_{ab}^{\sigma S}dr_{ab}^{\sigma}dr_{ab}^{\sigma}dr_{ab}^{S}dr_{ab}^{\sigma S}\nonumber
\\&&\times\exp\{N\sum ir_{ab}^{\sigma}[q_{ab}^{\sigma}-(\sigma^{a}+t_{1}'\phi^{a})(\sigma^{b}+t_{1}'\phi^{b})]\nonumber
\\&&+N\sum_{ab}ir_{ab}^{S}[q_{ab}^{S}-(S^{a}+t_{2}'\Phi^{a})(S^{b}+t_{2}'\Phi^b)]\nonumber
\\&&+N\sum_{ab}ir_{ab}^{\sigma S}[q_{ab}^{\sigma S}-(\sigma^{a}+t_{1}'\phi^{a})(S^{b}+t_{2}'\Phi^{b})]\nonumber
\\&&+\frac{\beta}{2}\sum_{ab}[\sum_{\mu\geq 2}(1+t_{1})q_{ab}^{\sigma}x_{\mu}^{a}x_{\mu}^{b}+(1+t_{2})y_{\mu}^{a}y_{\mu}^{b}q_{ab}^{S}\nonumber
\\&&+\sqrt{1+t_{1}}\sqrt{1+t_{2}}(q_{ab}^{\sigma S}x_{\mu}^{a}y_{\mu}^{b}+x_{\mu}^{b}y_{\mu}^{a}q_{ba}^{\sigma S})]\},\nonumber
\end{eqnarray}

where in the third equality we introduced Edward-Anderson variables and their auxiliary variables via the integral representation of delta function.

Now we look at the integrals in the non-condensed part:

\begin{eqnarray}
&&\int\prod_{a,b}Dx^{a}Dy^{b}\exp\{\frac{\beta}{2}\sum_{ab}[(1+t_{1})q_{ab}^{\sigma}x^{a}x^{b}+(1+t_{2})y^{a}y^{b}q_{ab}^{S}
\\&&+\sqrt{(1+t_{1})(1+t_{2})}(q_{ab}^{\sigma S}x^{a}y^{b}+x^{b}y^{a}q^{\sigma S}_{ba})]\}=\int\prod_{a,b}dx^{a}dy^{b}\nonumber
\\&&\times\exp\{\frac{1}{2}\sum_{ab}-\delta_{ab}x^{a}x^{b}-\delta_{ab}y^{a}y^{b}
+\beta[(1+t_{1})q_{ab}^{\sigma}x^{a}x^{b}+(1+t_{2})y^{a}y^{b}q_{ab}^{S}\nonumber
\\&&+\sqrt{1+t_{1}}\sqrt{1+t_{2}}(q_{ab}^{\sigma S}x_{\mu}^{a}y_{\mu}^{b}+x_{\mu}^{b}y_{\mu}^{a}q_{ba}^{\sigma S})]\}\
\\&&=\int \prod_{a,b}dx^{a}dy^{b}\exp(-\frac{1}{2}z^{T}[\mathbb{1}-\beta \hat{q})z],\nonumber
\end{eqnarray}

where $\hat{q}$ is a square matrix with dimension $2n\times 2n$ and the vector of dimension $2n$:

\begin{eqnarray}
&&\hat{q}\equiv
\begin{pmatrix}
  (1+t_{1})q^{\sigma} & \sqrt{(1+t_{1})(1+t_{2})}q^{\sigma S}\\ 
  & \\
  \sqrt{(1+t_{1})(1+t_{2})}(q^{\sigma S})^{T} & (1+t_{2})q^{S}
\end{pmatrix}\nonumber
\\&&z\equiv \begin{pmatrix}
  x_{1},  ...,  x_{n},
  y_{1},
  ...,
  y_{n}
\end{pmatrix}^T.
\end{eqnarray}

We have that

\begin{eqnarray}
\int \prod_{a}dz^{a}\exp[-\frac{1}{2}z(\mathbb{1}-\beta \hat{q})z]=[\det(\mathbb{1}-\beta \hat{q})]^{\frac{1}{2}}.
\end{eqnarray}

We rescale the auxiliary Edwards-Anderson variables and the magnetization variables:

\begin{eqnarray}
&&x_{1}\rightarrow\sqrt{\frac{\beta N}{1+t_{1}}}m_{1}^{a}; y_{1}\rightarrow\sqrt{\frac{\beta N}{1+t_{2}}}m_{2}^{a};
\\&&r_{ab}^{\rho}\rightarrow i\frac{\alpha\beta^2}{2}r_{ab}^{\rho} \forall \rho; h^{a}\rightarrow \sqrt{2\epsilon\beta N} h^{a}.\nonumber
\end{eqnarray}

With these changes, we get the following expression for the average of the replicated partition function:

\begin{eqnarray}
&&\langle Z^n\rangle=\int\prod_{a}dm_{1}^{a}dm_{2}^{a}dh^{a}\prod_{ab}dq_{ab}^{\sigma}dq_{ab}^{S}dq_{ab}^{\sigma S}dr_{ab}^{\sigma}dr_{ab}^{S}dr_{ab}^{\sigma S}
\\&&\times\exp\{-\frac{\beta N}{2}\sum_{a}[\frac{(m_{1}^{a})^2}{1+t_{1}}+\frac{(m_{1}^{a})^2}{1+t_{2}}+2\epsilon(h^{a})^2]-\log[\det(\mathbb{1}-\beta \hat{q})]^{\frac{p}{2}}\nonumber
\\&&-\frac{N\alpha\beta^2}{2}\sum_{ab}(r_{ab}^{\sigma}q_{ab}^{\sigma}+r_{ab}^{S}q_{ab}^{S}+r_{ab}^{\sigma S}q_{ab}^{\sigma S})\nonumber
\\&&+\log[\sum_{\sigma^{n},S^{n}}\int\prod_{i}\prod_{a}D\Phi_{i}^{a}D\phi_{i}^{b}\exp(\frac{\alpha\beta^2}{2}\sum_{ab}(r_{ab}^{\sigma}(\sigma_{i}^{a}+t'\phi_{i}^{a})(\sigma_{i}^{b}+t_{1}'\phi_{i}^{b})\nonumber
\\&&+r_{ab}^{S}(S_{i}^{a}+t_{2}'\Phi_{i}^{a})(S_{i}^{b}+t_{2}'\Phi_{i}^{b})+r_{ab}^{\sigma S}(\sigma_{i}^{a}+t'\phi_{i}^{a})(S_{i}^{b}+t_{2}'\Phi_{i}^{b}))\nonumber
\\&&+\beta\sum_{i}\sum_{a}\xi_{i}^{1}(m_{1}^{a}(\sigma_{i}^{a}+t_{1}'\phi_{i}^{a})+m_{2}^{a}(S_{i}^{a}+t_{2}'\Phi_{i}^{a}))+2\beta \epsilon\sum_{a}S_{i}\sigma_{i}^{a}h^{a})]\}.\nonumber
\end{eqnarray}

Now we apply the replica symmetry ansatz:

\begin{eqnarray}
&&\frac{1}{2n}\sum_{a}(\frac{(m_{1}^{a})^2}{1+t_{1}}+\frac{(m_{2}^{a})^2}{1+t_{2}}+2\epsilon(h^{a})^2)=\frac{m_{1}^2}{2+2t_{1}}+\frac{m_{2}^2}{2+2t_{2}}+\epsilon h^2,
\\&&\frac{\alpha\beta}{2n}\sum_{ab}r_{ab}^{\sigma}q_{ab}^{\sigma}=\frac{(\Delta^{\sigma}-1)(1+t_{2})}{2t_{2}}Q^{\sigma}+\frac{\alpha\beta}{2}r^{\sigma}(Q^{\sigma}-q^{\sigma}), \nonumber
\\&&\frac{\alpha\beta}{2n}\sum_{ab}r_{ab}^{S}q_{ab}^{S}=\frac{(\Delta^{S}-1)(1+t_{1})}{2t}Q^{S}+\frac{\alpha\beta}{2}r^{S}(Q^{S}-q^{S}), \nonumber
\\&&\frac{\alpha\beta}{2n}\sum_{ab}r_{ab}^{\sigma S}q_{ab}^{\sigma S}=\Delta^{\sigma S}Q^{\sigma S}+\frac{\alpha\beta r^{\sigma S}}{2}(Q^{\sigma S}-q^{\sigma S}),\nonumber
\end{eqnarray}

with

\begin{eqnarray}
&&\Delta^{\sigma}\equiv 1+\alpha\beta\frac{t_{1}}{1+t_{1}}(R^{\sigma}-r^{\sigma});\Delta^{S}\equiv 1+\alpha\beta\frac{t_{2}}{1+t_{2}}(R^{S}-r^{S});\nonumber
\\&&\Delta^{\sigma S}\equiv \frac{\alpha\beta}{2}(R^{\sigma S}-r^{\sigma S}).
\end{eqnarray}

The matrix $\hat{q}$ has 4 eigenvalues, their degeneracies and values are

\begin{eqnarray}
&&g_{1}=1,\lambda_{1}=\frac{1}{2}\{a+(n-1)b+c+(n-1)d
\\&&-[(a+(n-1)b+c+(n-1)d)^2\nonumber
\\&&-4((a+(n-1)b)(c+(n-1)d)-(e+(n-1)f)^2)]^{\frac{1}{2}}\},\nonumber
\\&&g_{2}=1,\lambda_{2}=\frac{1}{2}\{a+(n-1)b+c+(n-1)d\nonumber
\\&&+[(a+(n-1)b+c+(n-1)d)^2\nonumber
\\&&-4((a+(n-1)b)(c+(n-1)d)-(e+(n-1)f)^2)]^{\frac{1}{2}}\},\nonumber
\\&&g_{3}=n-1, \lambda_{3}=\frac{1}{2}\{a-b+c-d\nonumber
\\&&-\sqrt{(a-b+c-d)^2+4[(-a+b)(c-d)+(e-f)^2]}\},\nonumber
\\&&g_{4}=n-1, \lambda_{4}=\frac{1}{2}\{a-b+c-d\nonumber
\\&&+\sqrt{(a-b+c-d)^2+4[(-a+b)(c-d)+(e-f)^2]}\};\nonumber
\end{eqnarray}

where

\begin{eqnarray}
&&a=1-\beta Q^{\sigma}(1+t_{1}), \,\,\,b=-\beta q^{\sigma}(1+t_{1}),
\\&&c=1-\beta Q^{S}(1+t_{2}), \,\,\,d=-\beta q^{S}(1+t_{2})\nonumber
\\&&e=-\beta Q^{\sigma S}\sqrt{(1+t_{1})(1+t_{2})}, \,\,\,f=-\beta q^{\sigma S}\sqrt{(1+t_{1})(1+t_{2})}.\nonumber
\end{eqnarray}

The last step is to do two Hubbard-Stratonovich transformations on the quadratic variables, but first we need to massage to terms to get an useful formula:

\begin{eqnarray}
&&\sum_{ab}r^{\sigma}(\sigma^{a}+t_{1}'\phi^{a})(\sigma^{b}+t_{1}'\phi^{b})+\sum_{ab}r^{S}(S^{a}+t_{2}'\Phi^{a})(S^{b}+t_{2}')
\\&&+\sum_{ab}r^{\sigma S}(\sigma^{a}+t_{1}'\phi^{a})(S^{b}+t_{2}')=\frac{1}{2}[\sqrt{r^{\sigma}}\sum_{a} (\sigma^{a}+t_{1}'\phi^{a})\nonumber
\\&&+\sqrt{r^{S}}\sum_{a}(S^{a}+t_{2}'\Phi^{a})]^2(1+\frac{r^{\sigma S}}{2\sqrt{r^{\sigma}r^{S}}})+\frac{1}{2}[\sqrt{r^{\sigma}}\sum_{a}(\sigma^{a}+t_{1}'\phi^{a})\nonumber
\\&&-\sqrt{r^{S}}\sum_{a}(S^{a}+t_{2}'\Phi^{a})]^2(1-\frac{r^{\sigma S}}{2\sqrt{r^{\sigma}r^{S}}}).\nonumber
\end{eqnarray}

Only now   we  do the Hubbard-Stratonovich transformations in these quadratic terms. If we have an arbitrary number of agents $\aleph$, we can always do an analogous transformation, as it is equivalent to a change of coordinate system $\vec{x}\to\vec{x}'$ such that the $\sum_{i}^{\aleph}x_{i}^2+\sum_{i,j,i<j}^{\aleph}a_{i}x_{i}x_{j}=\sum_{i}^{\aleph} x_{i}^{' 2}$ with arbitrary coefficients $a_{i}$, but it is not hard to notice that this procedure leads to many additional terms, whose number scales faster than linear in  the number of agents. For comparison, the formula that should be used with 3 agents is the following:

\begin{eqnarray}
&&x^2+y^2+z^2+axy+bxz+cyz\nonumber
\\&&=\frac{1}{4}\{[(x+y)\sqrt{1+\frac{a}{2}}+z]^2(1+\frac{c+b}{2\sqrt{1+\frac{a}{2}}})\nonumber
\\&&+[(x+y)\sqrt{1+\frac{a}{2}}-z]^2(1-\frac{c+b}{2\sqrt{1+\frac{a}{2}}})\nonumber
\\&&+[(y-x)\sqrt{1-\frac{a}{2}}+z]^2(1+\frac{c-b}{2\sqrt{1-\frac{a}{2}}})\nonumber
\\&&+[(y-x)\sqrt{1-\frac{a}{2}}-z]^2(1-\frac{c-b}{2\sqrt{1-\frac{a}{2}}})\}
\end{eqnarray}

This is the main reason why we believe that it is not easy to generalize our results for interactions between 3,4, or more neural networks.

After that manipulation, we can sum over the spin states and get the free energy.

\section{Free energy expression and equations of state}

The  free energy for two interacting agents is

\begin{eqnarray}
&&f(m_{1},m_{2},h,Q^{\sigma},Q^{S},Q^{\sigma S},q^{\sigma},q^{S},q^{\sigma S},t_{1},t_{2},\beta,\alpha,\epsilon)= 
\\&&\frac{m_{1}^2}{2+2t_{1}}+\frac{m_{2}^2}{2+2t_{2}}+\epsilon h^2+\frac{(\Delta^{\sigma}-1)(1+t_{1})}{2t_{1}}Q^{\sigma}+\frac{\log[\Delta^{\sigma}\Delta^{S}-t_{1}'^{2}t_{2}'^{2}\beta^{2}(\Delta^{\sigma S})^{2}]}{2\beta}\nonumber
\\&&+\frac{\alpha\beta}{2}r^{\sigma}(Q^{\sigma}-q^{\sigma})+\frac{(\Delta^{S}-1)(1+t_{2})}{2t_{2}}Q^{S}+\frac{\alpha\beta}{2}r^{S}(Q^{S}-q^{S})\nonumber
\\&&+\Delta^{\sigma S}Q^{\sigma S}+\frac{\alpha\beta r^{\sigma S}}{2}(Q^{\sigma S}-q^{\sigma S})\nonumber
\\&&+\frac{\alpha}{2\beta}\log\{\frac{1}{2}[2+\beta(1+t_{1})(q^{\sigma}-Q^{\sigma})+\beta(1+t_{2})(q^{S}-Q^{S})\nonumber
\\&&-\beta(((1+t_{1})(q^{\sigma}-Q^{\sigma})-(1+t_{2})(q^{S}-Q^{S}))^2\nonumber
\\&&+4(1+t_{1})(1+t_{2})(-Q^{\sigma S}+ q^{\sigma S})^2)^{\frac{1}{2}}]\}\nonumber
\\&&+\frac{\alpha}{2\beta}\log\{\frac{1}{2}[2+\beta(1+t_{1})(q^{\sigma}-Q^{\sigma})+\beta(1+t_{2})(q^{S}-Q^{S})\nonumber
\\&&+\beta(((1+t_{1})(q^{\sigma}-Q^{\sigma})-(1+t_{2})(q^{S}-Q^{S}))^2\nonumber
\\&&+4(1+t_{1})(1+t_{2})(-Q^{\sigma S}+ q^{\sigma S})^2)^{\frac{1}{2}}]\}\nonumber
\\&&-\alpha\{q^{\sigma}(1+t_{1})[1-\beta(Q^{S}-q^{S})(1+t_{2})]\nonumber
\\&&+ q^{S}(1+t_{2})[1-\beta(Q^{\sigma}-q^{\sigma})(1+t_{1})]+\beta(1+t_{1})(1+t_{2})q^{\sigma S}(-q^{\sigma S}+Q^{\sigma S})\}\nonumber
\\&&\times\{2[1-\beta(1+t_{1})(Q^{\sigma}-q^{\sigma})][1-\beta(1+t_{2})(Q^{S}-q^{S})]\nonumber
\\&&-2\beta^{2}(1+t_{1})(1+t_{2})(-Q^{\sigma S}+q^{\sigma S})^2\}^{-1}\nonumber
\\&&-\frac{1}{\beta}\int DxDy\log(L_{1})\nonumber
\\&&-[\Delta^{S}t_{1}'^{2}\beta(\alpha r^{\sigma}+m_{1}^{2}+\frac{1}{\beta^{2}t_{1}'^{4}})+\Delta^{\sigma}t_{2}'^{2}\beta(\alpha r^{S}+m_{2}^{2}+\frac{1}{\beta^{2}t_{2}'^{4}})\nonumber
\\&&+2t_{1}'^{2}t_{2}'^{2}\beta^{2}\Delta^{\sigma S}(\frac{r^{\sigma S}\alpha}{2}+m_{1}m_{2})][2\Delta^{\sigma}\Delta^{S}-2t_{1}'^{2}t_{2}'^{2}\beta^{2}(\Delta^{\sigma S})^{2}]^{-1},\nonumber
\end{eqnarray}

where we use the auxiliary definitions

\begin{eqnarray}
&&r_{\pm}=\sqrt{1\pm\frac{r_{\sigma S}}{2\sqrt{r^{\sigma}r^{S}}}}, t_{1}'=i\sqrt{\frac{t_{1}}{\beta(t_{1}+1)}},t_{2}'=i\sqrt{\frac{t_{2}}{\beta(t_{2}+1)}},
\\&&\eta=2\epsilon h+\frac{\Delta^{\sigma S}}{\Delta^{\sigma}\Delta^{S}-t_{1}'^{2}t_{2}'^{2}\beta^{2}(\Delta^{\sigma S})^{2}} , \nonumber
\\&&\upsilon=\{\Delta^{S}[m_{1}+\sqrt{\frac{\alpha r^{\sigma}}{2}}(r_{+}x+r_{-}y)]+t_{2}'^{2}\beta\Delta^{\sigma S}[m_{2}+\sqrt{\frac{\alpha r^{S}}{2}}(r_{+}x-r_{-}y)]\}\nonumber
\\&&\times[\Delta^{\sigma}\Delta^{S}-t_{1}'^{2}t_{2}'^{2}\beta^{2}(\Delta^{\sigma S})^{2}]^{-1},\nonumber
\\&&\Upsilon=\{\Delta^{\sigma}[m_{2}+\sqrt{\frac{\alpha r^{S}}{2}}(r_{+}x-r_{-}y)]+t_{1}'^{2}\beta\Delta^{\sigma S}[m_{1}+\sqrt{\frac{\alpha r^{\sigma}}{2}}(r_{+}x+r_{-}y)]\},\nonumber
\\&&\times[\Delta^{\sigma}\Delta^{S}-t_{1}'^{2}t_{2}'^{2}\beta^{2}(\Delta^{\sigma S})^{2}]^{-1},\nonumber
\\&&L_{1}=\cosh[\beta(\upsilon+\Upsilon)]\exp(\beta\eta)+\cosh[\beta(\upsilon-\Upsilon)]\exp(-\beta\eta),\nonumber
\\&&L_{2}=\cosh[\beta(\upsilon+\Upsilon)]\exp(\beta\eta)-\cosh[\beta(\upsilon-\Upsilon)]\exp(-\beta\eta),\nonumber
\\&&L_{3}=\sinh[\beta(\upsilon+\Upsilon)]\exp(\beta\eta)+\sinh[\beta(\upsilon-\Upsilon)]\exp(-\beta\eta),\nonumber
\\&&L_{4}=\sinh[\beta(\upsilon+\Upsilon)]\exp(\beta\eta)-\sinh[\beta(\upsilon-\Upsilon)]\exp(-\beta\eta).\nonumber
\end{eqnarray}

Using the variational principle that the partial derivatives of the free energy must be 0 at the equilibrium point, we get 15 different equations for 15 different variables. We can simplify them by writing 6 of the variables as functions of the other 9, we call them former dependent and the latter independent. 

The equations of the 6 dependent variables are

\begin{eqnarray}
&&r^{\sigma}=\{q^{\sigma}(1+t_{1})^{2}[1-\beta(Q^{S}-q^{S})(1+t_{2})]^{2}
\\&&+\beta(Q^{\sigma S}-q^{\sigma S})(1+t_{1})^{2}(1+t_{2})[q^{\sigma S}(1-\beta(Q^{S}-q^{S})(1+t_{2}))\nonumber
\\&&+q^{S}(1+t_{2})\beta(Q^{\sigma S}-q^{\sigma S})]\}\nonumber
\\&&\times\{[1-\beta(Q^{\sigma}-q^{\sigma})(1+t_{1})][1-\beta(Q^{S}-q^{S})(1+t_{2})]\nonumber
\\&&-(Q^{\sigma S}-q^{\sigma S})^{2}(1+t_{1})(1+t_{2})\beta^{2}\}^{-2},\nonumber
\\\nonumber
\\\nonumber
\\&&\Delta^{\sigma}=1+\alpha t_{1}\{(1-\beta(Q^{S}-q^{S})(1+t_{2}))^{2}(1-\beta(Q^{\sigma}-q^{\sigma})(1+t_{1}))
\\&&-(1-\beta(Q^{S}-q^{S})(1+t_{2}))(Q^{\sigma S}-q^{\sigma S})^{2}\beta^{2}(1+t_{1})(1+t_{2})\}\nonumber
\\&&\times\{[1-\beta(Q^{\sigma}-q^{\sigma})(1+t_{1})][1-\beta(Q^{S}-q^{S})(1+t_{2})]\nonumber
\\&&-(Q^{\sigma S}-q^{\sigma S})^{2}(1+t_{1})(1+t_{2})\beta^{2}\}^{-2},\nonumber
\\\nonumber
\\\nonumber
\\&&r^{S}=(q^{S}(1+t_{2})^{2}(1-\beta(Q^{\sigma}-q^{\sigma})(1+t_{1}))^{2}
\\&&+\beta(Q^{\sigma S}-q^{\sigma S})(1+t_{1})(1+t_{2})^{2}(q^{\sigma S}(1-\beta(Q^{\sigma}-q^{\sigma})(1+t_{1}))\nonumber
\\&&+q^{\sigma}(1+t_{1})\beta(Q^{\sigma S}-q^{\sigma S})))\nonumber
\\&&\times\{[1-\beta(Q^{\sigma}-q^{\sigma})(1+t_{1})][1-\beta(Q^{S}-q^{S})(1+t_{2})]\nonumber
\\&&-(Q^{\sigma S}-q^{\sigma S})^{2}(1+t_{1})(1+t_{2})\beta^{2}\}^{-2},\nonumber
\\\nonumber
\\\nonumber
\\&&\Delta^{S}=1+\alpha t_{2}\{[1-\beta(Q^{\sigma}-q^{\sigma})(1+t_{1})]^{2}[1-\beta(Q^{S}-q^{S})(1+t_{2})]
\\&&-[1-\beta(Q^{\sigma}-q^{\sigma})(1+t_{1})](Q^{\sigma S}-q^{\sigma S})^{2}\beta^{2}(1+t_{1})(1+t_{2})\}\nonumber
\\&&\times\{[1-\beta(Q^{\sigma}-q^{\sigma})(1+t_{1})][1-\beta(Q^{S}-q^{S})(1+t_{2})]\nonumber
\\&&-(Q^{\sigma S}-q^{\sigma S})^{2}(1+t_{1})(1+t_{2})\beta^{2}\}^{-2},\nonumber
\\\nonumber
\\\nonumber
\\&&\Delta^{\sigma S}=\alpha\beta (-q^{\sigma S}+Q^{\sigma S})(1+t_{1})(1+t_{2})
\\&&\times\{2[1-\beta(Q^{\sigma}-q^{\sigma})(1+t_{1})][1-\beta(Q^{S}-q^{S})(1+t_{2})]\nonumber
\\&&-2\beta^{2}(-Q^{\sigma S}+q^{\sigma S})^{2}(1+t_{1})(1+t_{2})\}^{-1},\nonumber
\\\nonumber
\\\nonumber
\\&&r^{\sigma S}=(1+t_{1})(1+t_{2})\{q^{\sigma S}
\\&&\times[(1-\beta(Q^{\sigma}-q^{\sigma})(1+t_{1}))(1-\beta(Q^{S}-q^{S})(1+t_{2}))\nonumber
\\&&-\beta^{2}(Q^{\sigma S}-q^{\sigma S})^{2}(1+t_{1})(1+t_{2})]^{-1}\nonumber
\\&&+2[(-q^{\sigma S}+Q^{\sigma S})\beta(q^{\sigma}(1+t_{1})(1-\beta(Q^{S}-q^{S})(1+t_{2}))\nonumber
\\&&+q^{S}(1+t_{2})(1-\beta(Q^{\sigma}-q^{\sigma})(1+t_{1}))+q^{\sigma S}(1+t_{1})(1+t_{2})\beta(-q^{\sigma S}+Q^{\sigma S}))]\nonumber
\\&&\times[(1-\beta(Q^{\sigma}-q^{\sigma})(1+t_{1}))(1-\beta(Q^{S}-q^{S})(1+t_{2}))\nonumber
\\&&-\beta^{2}(Q^{\sigma S}-q^{\sigma S})^{2}(1+t_{1})(1+t_{2})^{2}]^{-2}\}.\nonumber
\end{eqnarray}

\pagebreak

The equations of the 9 independent variables are

\begin{eqnarray}
&&h=\int DxDy\frac{L_{2}}{L_{1}},
\\\nonumber
\\\nonumber
\\&&m_{1}(\frac{1}{1+t_{1}}-\frac{\Delta^{S}t_{1}'^{2}\beta}{\Delta^{\sigma}\Delta^{S}-t_{1}'^{2}t_{2}'^{2}\beta^{2}(\Delta^{\sigma S})^{2}})=\frac{m_{2}t_{1}'^{2}t_{2}'^{2}\beta^{2}\Delta^{\sigma S}}{\Delta^{\sigma}\Delta^{S}-t_{1}'^{2}t_{2}'^{2}\beta^{2}(\Delta^{\sigma S})^{2}}\nonumber
\\&&+\int DxDy\frac{1}{L_{1}}(\frac{\Delta^{S}L_{3}}{\Delta^{\sigma}\Delta^{S}-t_{1}'^{2}t_{2}'^{2}\beta^{2}(\Delta^{\sigma S})^{2}}+\frac{t_{1}'^{2}\beta\Delta^{\sigma S}L_{4}}{\Delta^{\sigma}\Delta^{S}-t_{1}'^{2}t_{2}'^{2}\beta^{2}(\Delta^{\sigma S})^{2}}),
\\\nonumber
\\\nonumber
\\&&m_{2}(\frac{1}{1+t_{2}}-\frac{\Delta^{\sigma}t_{2}'^{2}\beta}{\Delta^{\sigma}\Delta^{S}-t_{1}'^{2}t_{2}'^{2}\beta^{2}(\Delta^{\sigma S})^{2}})=\frac{m_{1}t_{1}'^{2}t_{2}'^{2}\beta^{2}\Delta^{\sigma S}}{\Delta^{\sigma}\Delta^{S}-t_{1}'^{2}t_{2}'^{2}\beta^{2}(\Delta^{\sigma S})^{2}}\nonumber
\\&&+\int DxDy\frac{1}{L_{1}}(\frac{\Delta^{\sigma S}t_{2}'^{2}\beta L_{3}}{\Delta^{\sigma}\Delta^{S}-t_{1}'^{2}t_{2}'^{2}\beta^{2}(\Delta^{\sigma S})^{2}}+\frac{\Delta^{\sigma}L_{4}}{\Delta^{\sigma}\Delta^{S}-t_{1}'^{2}t_{2}'^{2}\beta^{2}(\Delta^{\sigma S})^{2}}),
\\\nonumber
\\\nonumber
\\&&\frac{Q^{\sigma}(1+t_{1})}{2t_{1}}=
\\&&-\frac{\Delta^{S}}{2\beta\Delta^{\sigma}\Delta^{S}-2t_{1}'^{2}t_{2}'^{2}\beta^{3}(\Delta^{\sigma S})^{2}}-[(\Delta^{S})^{2}t_{1}'^{2}\beta(\alpha r^{\sigma}+m_{1}^{2}+\frac{1}{\beta^{2}t_{1}'^{4}})\nonumber
\\&&+t_{1}'^{2}t_{2}'^{4}\beta^{3}(\Delta^{\sigma S})^{2}(\alpha r^{S}+m_{2}^{2}+\frac{1}{\beta^{2}t_{2}'^{4}})+2t_{1}'^{2}t_{2}'^{2}\beta^{2}\Delta^{\sigma S}\Delta^{S}(\frac{r^{\sigma S}\alpha}{2}+m_{1}m_{2})]\nonumber
\\&&\times\{2[\Delta^{\sigma}\Delta^{S}-t_{1}'^{2}t_{2}'^{2}\beta^{2}(\Delta^{\sigma S})^{2}]^{2}\}^{-1}\nonumber
\\&&+\int DxDy\{L_{1}[\Delta^{\sigma}\Delta^{S}-t_{1}'^{2}t_{2}'^{2}\beta^{2}(\Delta^{\sigma S})^{2}]^{2}\}^{-1}\nonumber
\\&&\times\{[-(\Delta^{S})^{2}(m_{1}+\sqrt{\frac{\alpha r^{\sigma}}{2}}(r_{+}x+r_{-}y))\nonumber
\\&&-t_{2}'^{2}\beta\Delta^{S}\Delta^{\sigma S}(m_{2}+\sqrt{\frac{\alpha r^{S}}{2}}(r_{+}x-r_{-}y))]L_{3}\nonumber
\\&&-[t_{1}'^{2}t_{2}'^{2}\beta^{2}(\Delta^{\sigma S})^{2}(m_{2}+\sqrt{\frac{\alpha r^{S}}{2}}(r_{+}x-r_{-}y))\nonumber
\\&&+t_{1}'^{2}\beta\Delta^{S}\Delta^{\sigma S}(m_{1}+\sqrt{\frac{\alpha r^{\sigma}}{2}}(r_{+}x+r_{-}y))]L_{4}-\Delta^{S}\Delta^{\sigma S}L_{2}\},\nonumber
\\\nonumber
\\\nonumber
\\&&\frac{Q^{S}(1+t_{2})}{2t_{2}}=-\frac{\Delta^{\sigma}}{2\beta\Delta^{\sigma}\Delta^{S}-2t_{1}'^{2}t_{2}'^{2}\beta^{3}(\Delta^{\sigma S})^{2}}
\\&&-[(\Delta^{\sigma})^{2}t_{2}'^{2}\beta(\alpha r^{S}+m_{2}^{2}+\frac{1}{\beta^{2}t_{2}'^{4}})+t_{2}'^{2}t_{1}'^{4}\beta^{3}(\Delta^{\sigma S})^{2}(\alpha r^{\sigma}+m_{1}^{2}+\frac{1}{\beta^{2}t'^{4}})\nonumber
\\&&+2t_{1}'^{2}t_{2}'^{2}\beta^{2}\Delta^{\sigma S}\Delta^{\sigma}(\frac{r^{\sigma S}\alpha}{2}+m_{1}m_{2})]\{2[\Delta^{\sigma}\Delta^{S}-t_{1}'^{2}t_{2}'^{2}\beta^{2}(\Delta^{\sigma S})^{2}]^{2}\}^{-1}\nonumber
\\&&+\int DxDy\{L_{1}[\Delta^{\sigma}\Delta^{S}-t_{1}'^{2}t_{2}'^{2}\beta^{2}(\Delta^{\sigma S})^{2}]^{2}\}^{-1}\nonumber
\\&&\times\{[-(\Delta^{\sigma})^{2}(m_{2}+\sqrt{\frac{\alpha r^{S}}{2}}(r_{+}x-r_{-}y))\nonumber
\\&&-t_{1}'^{2}\beta\Delta^{\sigma}\Delta^{\sigma S}(m_{1}+\sqrt{\frac{\alpha r^{\sigma}}{2}}(r_{+}x+r_{-}y))]L_{4}\nonumber
\\&&-[t_{1}'^{2}t_{2}'^{2}\beta^{2}(\Delta^{\sigma S})^{2}(m_{1}+\sqrt{\frac{\alpha r^{\sigma}}{2}}(r_{+}x+r_{-}y))\nonumber
\\&&+t_{2}'^{2}\beta\Delta^{\sigma}\Delta^{\sigma S}(m_{2}+\sqrt{\frac{\alpha r^{S}}{2}}(r_{+}x-r_{-}y))]L_{3}-\Delta^{\sigma}\Delta^{\sigma S}L_{2}\},\nonumber
\\\nonumber
\end{eqnarray}

\begin{eqnarray}
&&Q^{\sigma S}=\frac{t_{1}'^{2}t_{2}'^{2}\beta\Delta^{\sigma S}}{\Delta^{\sigma}\Delta^{S}-t_{1}'^{2}t_{2}'^{2}\beta^{2}(\Delta^{\sigma S})^2}+\frac{t_{1}'^{2}t_{2}'^{2}\beta^2(\frac{r^{\sigma S}}{2}+m_{1}m_{2})}{\Delta^{\sigma}\Delta^{S}-t_{1}'^{2}t_{2}'^{2}\beta^{2}(\Delta^{\sigma S})^2}
\\&&+\{t_{1}'^{2}t_{2}'^{2}\beta^{2}\Delta^{\sigma S}[\Delta^{S}t_{1}'^{2}\beta(\alpha r^{\sigma}+m_{1}^{2}+\frac{1}{\beta^{2}t_{1}'^{4}})+\Delta^{\sigma}t_{2}'^{2}\beta(\alpha r^{S}+m_{2}^{2}+\frac{1}{\beta^{2}t_{2}'^{4}})\nonumber
\\&&+2t_{1}'^{2}t_{2}'^{2}\beta^{2}\Delta^{\sigma S}(\frac{r^{\sigma S}\alpha}{2}+m_{1}m_{2})]\}[\Delta^{\sigma}\Delta^{S}-t_{1}'^{2}t_{2}'^{2}\beta^{2}(\Delta^{\sigma S})^{2}]^{-2}\nonumber
\\&&+\int DxDy\{L_{1}[\Delta^{\sigma}\Delta^{S}-t_{1}'^{2}t_{2}'^{2}\beta^{2}(\Delta^{\sigma S})^{2}]^{2}\}^{-1}\{L_{3}\nonumber
\\&&\times[2t_{1}'^{2}t_{2}'^{2}\beta^{2}\Delta^{\sigma S}\Delta^{S}(m_{1}+\sqrt{\frac{\alpha r^{\sigma}}{2}}(r_{+}x+r_{-}y))+(t_{1}'^{2}t_{2}'^{4}\beta^{3}(\Delta^{\sigma S})^{2}+t_{2}'^{2}\beta\Delta^{\sigma}\Delta^{S})\nonumber
\\&&\times(m_{2}+\sqrt{\frac{\alpha r^{S}}{2}}(r_{+}x-r_{-}y))]+[2t_{1}'^{2}t_{2}'^{2}\beta^{2}\Delta^{\sigma S}\Delta^{\sigma}(m_{2}+\sqrt{\frac{\alpha r^{S}}{2}}(r_{+}x-r_{-}y))\nonumber
\\&&+(t_{1}'^{2}\beta\Delta^{\sigma}\Delta^{S}+t_{1}'^{4}t_{2}'^{2}\beta^{3}(\Delta^{\sigma S})^{2})(m_{1}+\sqrt{\frac{\alpha r^{\sigma}}{2}}(r_{+}x+r_{-}y))]L_{4}\nonumber
\\&&+[\Delta^{\sigma}\Delta^{S}+t_{1}'^{2}t_{2}'^{2}\beta^{2}(\Delta^{\sigma S})^{2}]L_{2}\},\nonumber
\\\nonumber
\\\nonumber
\\\nonumber
\\&&q^{\sigma}=Q^{\sigma}-\frac{\Delta^{S}t_{1}'^{2}}{\Delta^{\sigma}\Delta^{S}-t_{1}'^{2}t_{2}'^{2}\beta^{2}(\Delta^{\sigma S})^{2}}-\frac{1}{\alpha\beta}\int DxDy
\\&&\times\{L_{1}[4\Delta^{\sigma}\Delta^{S}-4t_{1}'^{2}t_{2}'^{2}\beta^{2}(\Delta^{\sigma S})^{2}]\}^{-1}\{L_{3}\nonumber
\\&&\times[\Delta^{S}(4\sqrt{\frac{\alpha}{2r^{\sigma}}}(r_{+}x+r_{-}y)\nonumber+\frac{r^{\sigma S}}{r^{\sigma}}\sqrt{\frac{\alpha}{2r^{S}}}(-\frac{x}{r_{+}}+\frac{y}{r_{-}}))\nonumber
\\&&-r^{\sigma S}t_{2}'^{2}\beta\Delta^{\sigma S}\sqrt{\frac{\alpha}{2(r^{\sigma})^{3}}}(\frac{x}{r_{+}}+\frac{y}{r_{-}})]\nonumber
\\&&+L_{4}[t_{1}'^{2}\beta\Delta^{\sigma S}(4\sqrt{\frac{\alpha}{2r^{\sigma}}}(r_{+}x+r_{-}y)+\frac{r^{\sigma S}}{r^{\sigma}}\sqrt{\frac{\alpha}{2r^{S}}}(-\frac{x}{r_{+}}+\frac{y}{r_{-}}))\nonumber
\\&&-r^{\sigma S}\Delta^{\sigma}\sqrt{\frac{\alpha}{2(r^{\sigma})^{3}}}(\frac{x}{r_{+}}+\frac{y}{r_{-}})]\},\nonumber
\end{eqnarray}

\begin{eqnarray}
&&q^{S}=Q^{S}-\frac{\Delta^{\sigma}t_{2}'^{2}}{\Delta^{\sigma}\Delta^{S}-t_{1}'^{2}t_{2}'^{2}\beta^{2}(\Delta^{\sigma S})^{2}}-\frac{1}{\alpha\beta}\int DxDy
\\&&\times\{L_{1}[4\Delta^{\sigma}\Delta^{S}-4t_{1}'^{2}t_{2}'^{2}\beta^{2}(\Delta^{\sigma S})^{2}]\}^{-1}\{L_{4}\nonumber
\\&&\times[\Delta^{\sigma}(4\sqrt{\frac{\alpha}{2r^{S}}}(r_{+}x-r_{-}y)-\frac{r^{\sigma S}}{r^{S}}\sqrt{\frac{\alpha}{2r^{\sigma}}}(\frac{x}{r_{+}}+\frac{y}{r_{-}})]\nonumber
\\&&+r^{\sigma S}t_{1}'^{2}\beta\Delta^{\sigma S}\sqrt{\frac{\alpha}{2(r^{S})^{3}}}(-\frac{x}{r_{+}}+\frac{y}{r_{-}})\}+L_{3}\nonumber
\\&&\times[t_{2}'^{2}\beta\Delta^{\sigma S}(4\sqrt{\frac{\alpha}{2r^{S}}}(r_{+}x-r_{-}y)\nonumber
\\&&-\frac{r^{\sigma S}}{r^{S}}\sqrt{\frac{\alpha}{2r^{\sigma}}}(\frac{x}{r_{+}}+\frac{y}{r_{-}}))+r^{\sigma S}\Delta^{S}\sqrt{\frac{\alpha}{2(r^{S})^{3}}}(-\frac{x}{r_{+}}+\frac{y}{r_{-}})]\},\nonumber
\\\nonumber
\\\nonumber
\\&&q^{\sigma S}=Q^{\sigma S}-\frac{t_{1}'^{2}t_{2}'^{2}\beta\Delta^{\sigma S}}{\Delta^{\sigma}\Delta^{S}-t_{1}'^{2}t_{2}'^{2}\beta^{2}(\Delta^{\sigma S})^{2}}
\\&&-\frac{1}{\alpha\beta}\int DxDy\{2L_{1}[\Delta^{\sigma}\Delta^{S}-t_{1}'^{2}t_{2}'^{2}\beta^{2}(\Delta^{\sigma S})^{2}]\}^{-1}\nonumber
\\&&\times\{L_{3}[\Delta^{S}\sqrt{\frac{\alpha}{2r^{S}}}(\frac{x}{r_{+}}-\frac{y}{r_{-}})+t_{2}'^{2}\beta\Delta^{\sigma S}\sqrt{\frac{\alpha}{2r^{\sigma}}}(\frac{x}{r_{+}}+\frac{y}{r_{-}})]\nonumber
\\&&+L_{4}[\Delta^{\sigma}\sqrt{\frac{\alpha}{2r^{\sigma}}}(\frac{x}{r_{+}}+\frac{y}{r_{-}})+t_{1}'^{2}\beta\Delta^{\sigma S}\sqrt{\frac{\alpha}{2r^{S}}}(\frac{x}{r_{+}}-\frac{y}{r_{-}})]\}.\nonumber
\end{eqnarray}

\section{Zero temperature equations}

To obtain zero temperature equations, it is necessary to deal with the three following combinations:

\begin{eqnarray}
I_{1}=\frac{L_{2}}{L_{1}}, \,\,\,I_{2}=\frac{L_{3}}{L_{1}}, \,\,\, I_{3}=\frac{L_{4}}{L_{1}}.
\end{eqnarray}

They become 

\begin{eqnarray}
&&I_{1}=\frac{1}{4}\{[\sign(\upsilon+\Upsilon)+1][\sign(\upsilon-\Upsilon)+1]\sign(\eta+\Upsilon)
\\&&+[\sign(\upsilon+\Upsilon)+1][-\sign(\upsilon-\Upsilon)+1]\sign(\eta+\upsilon)\nonumber
\\&&+[-\sign(\upsilon+\Upsilon)+1][\sign(\upsilon-\Upsilon)+1]\sign(\eta-\upsilon)\nonumber
\\&&+[-\sign(\upsilon+\Upsilon)+1][-\sign(\upsilon-\Upsilon)+1]\sign(\eta-\Upsilon)\}.\nonumber
\\&&I_{2}=\frac{1}{4}\{[\sign(\upsilon+\Upsilon)+1][\sign(\upsilon-\Upsilon)+1]\nonumber
\\&&+[\sign(\upsilon+\Upsilon)+1][-\sign(\upsilon-\Upsilon)+1]\sign(\eta+\upsilon)\nonumber
\\&&+[-\sign(\upsilon+\Upsilon)+1][\sign(\upsilon-\Upsilon)+1]\sign(\upsilon-\eta)\nonumber
\\&&-[-\sign(\upsilon+\Upsilon)+1][-\sign(\upsilon-\Upsilon)+1]\}\nonumber
\\&&I_{3}=\frac{1}{4}\{[\sign(\upsilon+\Upsilon)+1][\sign(\upsilon-\Upsilon)+1]\sign(\eta+\Upsilon)\nonumber
\\&&+[-\sign(\upsilon+\Upsilon)+1][-\sign(\upsilon-\Upsilon)+1]\sign(\Upsilon-\eta)\nonumber
\\&&+[\sign(\upsilon+\Upsilon)+1][-\sign(\upsilon-\Upsilon)+1]\nonumber
\\&&-[-\sign(\upsilon+\Upsilon)+1][\sign(\upsilon-\Upsilon)+1]\}.\nonumber
\end{eqnarray}

It is possible to remove one of the integrals as it is done in \cite{AGS85} and \cite{Fachechi2019}, but in this case the equations would become substantially bigger and there would not be a visible simplification. Besides that change, it is necessary to substitute $q^{\rho}$ by $c^{\rho}\equiv\beta(Q^{\rho}-q^{\rho})$ in the same way that it is done in \cite{AGS85} and \cite{Fachechi2019}. We did not find any particularly interesting property in this region, so we did not focus on it.


\end{document}